\documentstyle[12pt,equation,axodraw]{article}
\begin{document}
\newcommand{\be}{\begin{equation}}
\newcommand{\ben}{\begin{subequations}}
\newcommand{\een}{\end{subequations}}
\newcommand{\beq}{\begin{eqalignno}}
\newcommand{\eeq}{\end{eqalignno}}
\newcommand{\ee}{\end{equation}}
\newcommand{\mchi}{\mbox{$m_\chi$}}
\newcommand{\Ochi}{\mbox{$\Omega_\chi h^2$}}
\newcommand{\tanb}{\mbox{$\tan \! \beta$}}
\newcommand{\cotb}{\mbox{$\cot \! \beta$}}
\newcommand{\cosb}{\mbox{$\cos \! \beta$}}
\newcommand{\sinb}{\mbox{$\sin \! \beta$}}
\newcommand{\stw}{\mbox{$\sin^2 \! \theta_W$}}
\newcommand{\ksl}{k \hspace*{-2.2mm} /}
\newcommand{\mpl}{\mbox{$M_{Pl}$}}
\newcommand{\mx}{\mbox{$M_X$}}
\newcommand{\sym}{\mbox{$SU(2) \times U(1)_Y$}}
\newcommand{\ddk}{\frac{d^4 k}{(2\pi)^4}}
\newcommand{\sfl}{\mbox{$\tilde{f}_L$}}
\newcommand{\sfr}{\mbox{$\tilde{f}_R$}}
\newcommand{\sff}{\mbox{$\tilde{f}$}}
\newcommand{\ltf}{\mbox{$\tilde{\lambda}_f$}}
\newcommand{\thb}{\mbox{$\overline{\theta}$}}
\newcommand{\alb}{\mbox{$\overline{\alpha}$}}
\newcommand{\qbar}{\mbox{$\overline{Q}$}}
\newcommand{\lag}{\mbox{${\cal L}$}}
\newcommand{\ter}{\mbox{${\tilde{e}_R^c}$}}
\newcommand{\tur}{\mbox{${\tilde{u}_R^c}$}}
\newcommand{\tdr}{\mbox{${\tilde{d}_R^c}$}}
\newcommand{\tll}{\mbox{${\tilde{l}_L}$}}
\newcommand{\tql}{\mbox{${\tilde{q}_L}$}}
\newcommand{\ttl}{\mbox{${\tilde{t}_L}$}}
\newcommand{\ttr}{\mbox{${\tilde{t}_R}$}}
\newcommand{\tbl}{\mbox{${\tilde{b}_L}$}}
\newcommand{\tbr}{\mbox{${\tilde{b}_R}$}}
\newcommand{\tg}{\mbox{${\tilde{g}}$}}
\newcommand{\tc}{\mbox{${\tilde{\chi}}$}}
\newcommand{\dtm}{\mbox{${\cos(2\beta) M_Z^2}$}}
\newcommand{\piff}{\pi_{\phi\phi}}
\newcommand{\wt}{\widetilde}

\pagestyle{empty}

\begin{flushright}
APCTP--5 \\
KEK--TH--501 \\
November 1996\\
\end{flushright}
\vspace*{1.5cm}

\begin{center}
{\Large \bf An Introduction to Supersymmetry} \\
\vspace{1cm}
{\large Manuel Drees} \\
{\it Asia--Pacific Center for Theoretical Physics, Seoul, Korea}\\
\end{center}

\begin{abstract}
A fairly elementary introduction to supersymmetric field theories in
general and the minimal supersymmetric Standard Model (MSSM) in particular
is given. Topics covered include the cancellation of quadratic divergencies,
the construction of the supersymmetric Lagrangian using superfields, the
field content of the MSSM, electroweak symmetry breaking in the MSSM,
mixing between different superparticles (current eigenstates) to produce
mass eigenstates, and the embedding of the MSSM in so--called minimal
supergravity.
\end{abstract}
\clearpage
\setcounter{page}{1}
\pagestyle{plain}

\section*{1. Introduction}
In the last 20 years the SLAC Spires data base has registered almost
10,000 papers dealing with various aspects of supersymmetric field theories.
This is quite remarkable, given that there is {\em no} direct experimental
evidence for the existence of any of the numerous new particles predicted
by such theories. This apparent discrepancy between theoretical speculation
and experimental fact has even caught the public eye, and led to charges
that modern particle physics resembles medieval alchemy.

I will therefore start these lecture notes by reviewing in some detail
the main argument for the existence of supersymmetric particles ``at
the weak scale'' (i.e., with mass very roughly comparable to those of
the heaviest known elementary particles, the $W$ and $Z$ bosons and
the top quark). This argument rests on the observation that
supersymmetric field theories ``naturally'' allow to chose the weak
scale to be many orders of magnitude below the hypothetical scale \mx\
of Grand Unification or the Planck scale \mpl. This is closely related to
the cancellation of quadratic divergencies \cite{2} in supsersymmetric field
theories; such divergencies are notorious in non--supersymmetric
theories with elementary scalar particles, such as the Standard Model
(SM). In Sec.~2 this question will be discussed in more detail, and
the cancellation of quadratic divergencies involving Yukawa interactions
will be demonstrated explicitly (in 1--loop order).

This explicit calculation will indicate the basic features that the 
proposed new symmetry has to have if it is to solve the ``naturalness
problem'' \cite{3} of the SM. In particular, we will need equal numbers
of physical (propagating) bosonic and fermionic degrees of freedom;
also, certain relations between the coefficients of various terms in the
Lagrangian will have to hold. In Sec.~3 we will discuss a method that allows
to quite easily construct field theories that satisfy these conditions,
using the language of superfields. This will be the most formal part of
these notes. At the end of this section, the Lagrangian will have been
constructed, and we will be ready to check the cancellation of
quadratic divergencies due to gauge interactions. This involves a far
greater number of diagrams and fields than the case of Yukawa
interactions; it seems quite unlikely that one could have hit on the
necessary set of fields and their interactions using the kind of
guesswork that will be used (with hindsight) in Sec.~2. At the end
of Sec.~3 the problem of supersymmetry breaking will be discussed
briefly.

Having hopefully convinced the reader that supersymmetric field theories
are interesting, and having shown how to construct them in general, in Sec.~4
I attempt to make contact with reality by discussing several issues
related to the phenomenology of the simplest potentially realistic
supersymmetric field theory, the Minimal Supersymmetric Standard Model
or MSSM. I will begin this section with a review of the motivation for
considering a supersymmetrization of the SM. The absence of quadratic
divergencies remains the main argument, but the MSSM also has several
other nice features not shared by the SM. In Sec.~4a the field content
of the model will be listed, and the Lagrangian will be written down;
this is an obvious application of the results of Sec.~3. In Sec.~4b
the breaking of the electroweak gauge symmetry will be discussed. This
plays a central role both theoretically (since without elementary scalar
``Higgs'' bosons the main argument for weak--scale supersymmetry collapses)
and phenomenologically (since it will lead to a firm and, at least in
principle, easily testable prediction). Next, mixing between various
superparticles (``sparticles'') will be discussed. This mixing, which is
a direct consequence of \sym\ gauge symmetry breaking, unfortunately makes
the correspondence between particles and sparticles less transparent.
However, an understandig of sparticle mixing is essential for an 
understanding of almost all ongoing work on the phenomenology of the MSSM.

The least understood aspect of the MSSM concerns the breaking of supersymmetry.
A general parametrization of this (necessary) phenomenon introduces more than
100 free parameters in the model. Fortunately not all of these parameters
will be relevant for a given problem or process, at least not in leading
order in perturbation theory. Nevertheless, it is of interest to look for
schemes that attempt to reduce the number of free parameters. The most
popular such scheme is (loosely) based on the extension of global
supersymmetry to its local version, supergravity, and is hence known
as ``minimal supergravity'' or mSUGRA. This model is attractive not
only because of its economy and resulting predictive power, but also
because it leads to a dynamical explanation (as opposed to a mere
parametrization) of electroweak symmetry breaking. This will be discussed
in Sec.~4d. I will in conclude Sec.~5 by briefly mentioning some areas of
active research. 

\section*{2. Quadratic Divergencies}
This section deals with the problem of quadratic divergencies in the SM,
and an explicit calculation is performed to illustrate how the introduction
of new fields with judicioulsy chosen couplings can solve this problem. In
order to appreciate the ``bad'' quantum behaviour of the scalar sector of
the SM, let us first briefly review some corrections in QED, the best
understood ingredient of the SM.

The examples studied will all be two--point functions (inverse
propagators) at vanishing external momentum, computed at one--loop
level. The calculations will therefore be quite simple, yet they
suffice to illustrate the problem. Roughly speaking, the computed
quantity corresponds to the mass parameters appearing in the Lagrangian;
since I will assume vanishing external momentum, this will not be the
on--shell (pole) mass, but it is easy to see that the difference between
these two quantities can at most involve logarithmic divergencies (due
to wave function renormalization).

\begin{center}
\begin{picture}(300,100)(0,0)
\Photon(75,50)(125,50){4}{5.5}
\Photon(175,50)(225,50){4}{5.5}
\ArrowArc(150,50)(25,0,180)
\ArrowArc(150,50)(25,180,360)
\Text(150,20)[]{$e^-$}
\Text(150,85)[]{$e^+$}
\Text(65,50)[]{$\gamma$}
\Text(235,50)[]{$\gamma$}
\end{picture}

{\bf Fig.~1:} The photon self--energy diagram in QED.
\end{center}

Let us first investigate the photon's two--point function, which
receives contributions due to the electron loop diagram of Fig.~1:
\begin{eqnarray} \label{e1}
\pi_{\gamma \gamma}^{\mu \nu} (0) &=& - \int \ddk {\rm tr} \left[ \left(
- i e \gamma^{\mu} \right) \frac {i} {\ksl - m_e} \left( - i e
\gamma^{\nu} \right) \frac {i} {\ksl - m_e} \right] \nonumber \\
&=& -4 e^2 \int \ddk \frac {2 k^{\mu} k^{\nu} - g^{\mu\nu} \left( k^2
- m_e^2 \right) } { \left( k^2 - m_e^2 \right)^2 } \nonumber \\
&=& 0.
\end{eqnarray}
The fact that the integral in eq.(\ref{e1}) vanishes is manifest only
in a regularization scheme that preserves gauge invariance, e.g.
dimensional regularization. On a deeper level, this result is the
consequence of the exact $U(1)$ gauge invariance of QED, which ensures
that the photon remains massless in all orders in perturbation theory.

\begin{center}
\begin{picture}(300,100)(0,0)
\ArrowLine(75,50)(225,50)
\PhotonArc(150,50)(25,0,180){4}{8.5}
\Text(150,85)[]{$\gamma$}
\Text(65,50)[]{$e^-$}
\end{picture}

{\bf Fig.~2:} The electron self energy in QED.
\end{center}

Next, let us consider the electron self energy correction of Fig.~2:
\begin{eqnarray} \label{e2}
\pi_{ee}(0) &=& \int \ddk \left( -i e \gamma_{\mu} \right) \frac {i}
{\ksl - m_e} \left( - i e \gamma_{\nu} \right) \frac {-i g^{\mu\nu}} {k^2}
\nonumber \\
&=& - e^2 \int \ddk \frac {1} {k^2 \left( k^2 - m_e^2 \right)} 
\gamma_{\mu} \left( \ksl + m_e \right) \gamma^{\mu} \nonumber \\
&=& - 4 e^2 m_e \int \ddk \frac {1} {k^2 \left( k^2 - m_e^2 \right)}.
\end{eqnarray}
In the last step I have made use of the fact that the $\ksl-$term
in the numerator vanishes after integration over angles, if one 
uses a regulator that respects Poincar\'e invariance. The integral
in eq.(\ref{e2}) has a logarithmic divergence in the ultraviolet (at
large momenta). Notice, however, that this correction to the electron
mass is itself proportional to the electron mass. The coefficient is
formally infinite; however, even if we replace the ``infinity'' by
the largest scale in particle physics, the Planck scale, we find a correction
\be \label{e3}
\delta m_e \simeq 2 \frac {\alpha_{\rm em}} {\pi} m_e \log \frac {\mpl}
{m_e} \simeq 0.24 m_e,
\ee
which is quite modest. At a deeper level, the fact that this correction is
quite benign can again be understood from a symmetry: In the limit $m_e
\rightarrow 0$, the model becomes invariant under chiral rotations
$\psi_e \rightarrow \exp(i \gamma_5 \varphi) \psi_e$. If this symmetry were
exact, the correction of eq.(\ref{e2}) would have to vanish. In reality
the symmetry is broken by the electron mass, so the correction must
itself be proportional to $m_e$.

\begin{center}
\begin{picture}(300,100)(0,0)
\DashLine(75,50)(125,50){4}
\DashLine(175,50)(225,50){4}
\ArrowArc(150,50)(25,0,180)
\ArrowArc(150,50)(25,180,360)
\Text(150,15)[]{$f$}
\Text(150,85)[]{$\bar{f}$}
\Text(65,50)[]{$\phi$}
\Text(235,50)[]{$\phi$}
\end{picture}

{\bf Fig.~3:} A fermion anti--fermion contribution to the self energy of the
Higgs boson in the Standard Model.
\end{center}

Now consider the contribution of heavy fermion loops to the two--point
function of the SM Higgs field $\phi = \Re (H-v)/\sqrt{2}$, shown in
Fig.~3. Let the
$H f \bar{f}$ coupling be $\lambda_f$; the correction is then given by
\begin{eqnarray} \label{e4}
\pi_{\phi\phi}^{f} (0) &=& - N(f) \int \ddk {\rm tr} \left[ \left(
i \frac{\lambda_f} {\sqrt{2}} \right) \frac {i} {\ksl - m_f} \left( i
\frac{\lambda_f} {\sqrt{2}} \right) \frac {i} {\ksl - m_f} \right]
\nonumber \\
&=& - 2 N(f) \lambda_f^2 \int \ddk \frac {k^2 + m_f^2} {\left( k^2 - m_f^2
\right)^2 } \nonumber \\
&=& - 2 N(f) \lambda_f^2 \int \ddk \left[ \frac {1}{k^2 - m_f^2}
+ \frac {2 m_f^2} {\left( k^2 - m_f^2 \right)^2} \right].
\end{eqnarray}
Here, $N(f)$ is a multiplicity factor (e.g., $N(t)=3$ for top quarks, due
to summation over color indices). 

The first term in the last line of eq.(\ref{e4}) is {\em
quadratically} divergent! If we were to replace the divergence
$\Lambda^2$ by $M^2_{Pl}$, the resulting ``correction'' would be some
30 orders of magnitude {\em larger} than the physical SM Higgs mass,
$m_{\phi} \leq 1$ TeV (to preserve unitarity of $WW$ scattering
amplitudes \cite{3a}). The contrast to the modest size of the
correction (\ref{e3}) dramatically illustrates the difference between
logarithmic and quadratic divergencies. Note also that the correction
(\ref{e4}) is itself independent of $m_{\phi}$. This is related to the
fact that setting $m_{\phi} = 0$ does {\em not} increase the symmetry
group of the SM. There is nothing in the SM that ``protects'' the
Higgs mass in the way that the photon and even the electron masses are
protected.

Of course, one can still simply renormalize the quadratic divergence away,
as one does with logarithmic divergencies. However, I just argued that
logarithmic and quadratic divergencies are indeed quite different. Besides, 
this would still leave us with a finite correction from eq.(\ref{e4}), of
order $N(f) m_f^2 \lambda_f^2 / 8 \pi$. Such a correction would be quite
small if $f$ is an SM fermion like the top quark. However, it is quite
unlikely that the SM is indeed the ``ultimate theory''; it is much more
plausible that at some very high energy scale it will have to be replaced
by some more fundamental theory, for example a Grand Unified model \cite{4}.
In this case there will be corrections like those of eq.(\ref{e4}), with
$m_f$ being of order of this new (very high) scale. One would then need
extreme finetuning to cancel a very large bare mass against very large
loop corrections, leaving a result of order 1 TeV or less. Moreover, the
finetuning would be very different in different orders of perturbation
theory. This is the (technical aspect of) the ``hierarchy problem'':
Scalar masses ``like'' to be close to the highest mass scale in the 
theory \cite{5}.

\begin{center}
\begin{picture}(400,100)(0,0)
\DashLine(20,50)(180,50){4}
\DashCArc(100,75)(25,0,360){4}
\DashLine(220,50)(270,50){4}
\DashLine(320,50)(370,50){4}
\DashCArc(295,50)(25,0,360){4}
\Text(10,50)[]{$\phi$}
\Text(65,75)[]{$\tilde{f}$}
\Text(210,50)[]{$\phi$}
\Text(380,50)[]{$\phi$}
\Text(295,15)[]{$\tilde{f}$}
\Text(295,85)[]{$\tilde{f}$}
\end{picture}

{\bf Fig.~4:} Sfermion loop contributions to the Higgs self energy. $\tilde{f}$
stands for either $\tilde{f}_L$ or $\tilde{f}_R$.
\end{center}

In supersymmetric field theories this problem is solved since there are
additional contributions to $\pi_{\phi\phi}$. For example, let us
introduce two complex scalar fields \sfl, \sfr, with the following coupling
to the Higgs field:
\be \label{e5}
{\cal L}_{\phi \tilde{f}} = \frac{1}{2} \ltf \phi^2 \left( \left|
\sfl \right|^2 + \left| \sfr \right|^2 \right) + v \ltf \phi 
\left( \left| \sfl \right|^2 + \left| \sfr \right|^2 \right) +
\left( \frac {\lambda_f}{\sqrt{2}} A_f \phi \sfl \sfr^{\ast} + h.c. \right).
\ee
Here, $v$ is the vacuum expectation value of the SM Higgs field
($v \simeq 246$ GeV). The second term in the Lagrangian (\ref{e5}) is
thus due to the breaking of \sym, and its coefficient is related to
that of the first term. The coefficient of the third term is left
arbitrary, however. (The factor $\lambda_f$ appears here just by
convention.) This Lagrangian gives the following contribution to 
$\pi_{\phi\phi}$ (I assume the multiplicity factor $N$ to be the same
for \sfl\ and \sfr):
\begin{eqnarray} \label{e6}
\pi_{\phi\phi}^{\tilde{f}} (0) &=& - \ltf N(\sff) \int \ddk \left[
\frac {1} {k^2 - m^2_{\tilde{f}_L} } + \frac {1}
{k^2 - m^2_{\tilde{f}_R}} \right] \nonumber \\
&+& \left( \ltf v \right)^2 N(\sff) \int \ddk \left[ \frac {1}
{ \left( k^2 - m^2_{\tilde{f}_L} \right)^2 } +
\frac {1} { \left( k^2 - m^2_{\tilde{f}_R} \right)^2 }
\right] \nonumber \\
&+& \left| \lambda_f A_f \right|^2 N(\sff) \int \ddk  \frac {1}
{ \left( k^2 - m^2_{\tilde{f}_L} \right) 
\left( k^2 - m^2_{\tilde{f}_R} \right) }
\end{eqnarray}
Only the first line in eq.(\ref{e6}), which comes from the left diagram in
Fig.~4, contains quadratically divergent
terms. Comparing this with the fermionic contribution of eq.(\ref{e4}),
we see that the quadratic divergencies can be made to cancel by choosing
\ben \label{e7} \beq
N(\sfl) &= N(\sfr) = N(f); \label{e7a} \\
\ltf &= - \lambda_f^2 . \label{e7b}
\eeq \een
(Note that $\ltf<0$ is required for the scalar potential to be bounded
from below.)

Notice that the cancellation of quadratic divergencies does not impose any
restrictions on the masses $m_{\tilde{f}_L}, \ m_{\tilde{f}_R}$, nor
on the coupling $A_f$. Let us now sum eqs.(\ref{e4}) and (\ref{e6}) after
eqs.(\ref{e7}) have been imposed. To this end, I will use $\overline{MS}$
(or $\overline{DR}$ \cite{6}) regularization for all divergencies, which
gives:
\ben \label{e8} \beq
\int \frac {d^4 k} {i \pi^2} \frac {1} {k^2 - m^2} &= m^2 \left( 1 - \log
\frac {m^2} {\mu^2} \right) ; \label{e8a} \\
\int \frac {d^4 k} {i \pi^2} \frac {1} {\left(k^2 - m^2 \right)^2} &= 
-\log \frac {m^2} {\mu^2} , \label{e8b} 
\eeq \een
where $\mu$ is the renormalization scale.
For simplicity, let us assume $m_{\tilde{f}_L} = m_{\tilde{f}_R}
\equiv m_{\tilde{f}}$; the result can easily be generalized. This
gives:
\begin{eqnarray} \label{e9}
\pi_{\phi\phi}^{f+\tilde{f}} (0) = i \frac {\lambda^2_f N(f)}
{16 \pi^2} & {} & \left[ - 2 m^2_f \left( 1 - \log \frac {m^2_f} {\mu^2}
\right) + 4 m_f^2 \log \frac {m_f^2} {\mu^2} \right. \nonumber \\ 
&{}& \left. + 2 m^2_{\tilde{f}} \left( 1 - \log \frac {m^2_{\tilde{f}}}
{\mu^2} \right) - 4 m^2_f \log \frac {m^2_{\tilde{f}}} {\mu^2} 
\right. \nonumber \\
&{}& \left. - |A_f|^2 \log \frac {m^2_{\tilde{f}}} {\mu^2} \right].
\end{eqnarray}
Here, the first line is the fermionic contribution of eq.(\ref{e4}),
and the next three terms corresponds to the three lines of 
eq.(\ref{e6}); in the next--to--last term, I have used the relation
$m_f = \lambda_f v / \sqrt{2}$, which is true for SM fermions. From
eq.(\ref{e9}) we see that we can achieve a complete cancellation
between the fermionic and bosonic contributions, i.e. a {\em
vanishing} total correction, if we require in addition to eqs.(\ref{e8}):
\ben \label{e10} \beq
m_{\tilde{f}} &= m_f; \label{e10a} \\
A_f &= 0. \label{e10b}
\eeq \een
The fact that this leads to a vanishing total correction strongly hints at
the existence of an additional symmetry, as the discussion of QED 
radiative corrections at the beginning of this section shows. This line
of reasoning will be pursued in the next section. Before we turn to this,
let us see what happens if we violate the conditions (\ref{e10}) only
a little, i.e. if we take $m^2_{\tilde{f}} = m^2_f + \delta^2$, with
$\delta, \ |A_f| \ll m_f$, so that $\log \frac {m^2_{\tilde{f}}}
{\mu^2} \simeq \log \frac {m_f^2} {\mu^2} + \frac {\delta^2} {m_f^2}$:
\begin{eqnarray} \label{e11}
\pi_{\phi\phi}^{f+\tilde{f}} &\simeq & i \frac {\lambda_f^2 N(f)}
{16 \pi^2} \left[ - 2 \delta^2 \log \frac {m_f^2}{\mu^2} - 4 \delta^2
- |A_f|^2 \log \frac {m_f^2}{\mu^2} \right] + {\cal O}(\delta^4, A_f^2
\delta^2) \nonumber \\
&=& - i \frac {\lambda_f^2 N(f)} {16 \pi^2} \left[ 4 \delta^2 +
\left( 2 \delta^2 + |A_f|^2 \right) \log \frac {m_f^2} {\mu^2} 
\right] + {\cal O}(\delta^4, A_f^2 \delta^2).
\end{eqnarray}
(The first term in the first line of eq.(\ref{e10}) comes from the
first and third terms in eq.(\ref{e9}), and the second term from the
second and fourth terms.) We thus find the remarkable result that even
if we send $m_f$ to infinity, the correction (\ref{e11}) will remain
of modest size as long as the {\em difference} between $m^2_f$ and
$m^2_{\tilde{f}}$, as well as the coefficient $|A_f|$, remain small.
Thus the introduction of the fields \sfl\ and \sfr\ has not only allowed 
us to cancel quadratic divergencies; it also shields the weak scale
from loop corrections involving very heavy particles, {\em provided}
the mass splitting between fermions and bosons is itself of the weak
scale.\footnote{The alert reader will have noticed that I cheated a
little. In the derivation of eq.(\ref{e9}) I used $m_f = \lambda_f v
/ \sqrt{2}$. Sending $m_f \rightarrow \infty$ then amounts to to
sending $\lambda_f \rightarrow \infty$, if $v$ is fixed to its SM
value; in this case the correction (\ref{e11}) is still large, and
perturbation theory becomes altogether unreliable. However, the main
result survives in a more careful treatment of models with two
very different mass scales: The low scale is shielded from the high
scale as long as the mass splitting between bosons and fermions is
itself only of the order of the low scale. This already follows from
dimensional considerations, once we have shown that the corrections
vanish entirely in the limit where eqs.(\ref{e10}) hold exactly.}

\section*{3. Construction of Supersymmetric Field Theories}
This section describes how to construct the Lagrangian of a
supersymmetric field theory. To that end I first give a formal
definition of the supersymmetry (SUSY) algebra. The next two
subsections introduce chiral and vector superfields, respectively. The
construction of a Lagrangian with exact supersymmetry will be
accomplished in Sec.~3d. In the following subsection it will be shown
that this Lagrangian also leads to the cancellation of quadratic
divergencies from one--loop gauge contributions to the Higgs
two--point function $\piff(0)$, thereby extending the result
of the previous section. Finally, soft SUSY breaking is treated in
Sec.~3f.

Many excellent reviews of and introductions to the material covered here
already exist \cite{6,6a}; I will therefore be quite brief. My notation
will mostly follow that of Nilles \cite{6a}.

\subsection*{3a. The SUSY Algebra}
We saw in Sec.~2 how contributions to the Higgs two--point function
$\pi_{\phi\phi}(0)$ coming from the known SM fermions can be cancelled
exactly, if we introduce new bosonic fields with judiciously chosen
couplings. This strongly indicates that a new symmetry is at work here,
which can protect the Higgs mass from large (quadratically divergent)
radiative corrections, something that the SM is unable to 
do.\footnote{In principle one can cancel the one--loop quadratic divergencies
in the SM without introducing new fields, by explicitly cancelling
bosonic and fermionic contributions; this leads to a relation
between the Higgs and top masses \cite{7}. However, such a cancellation
would be purely ``accidental'', not enforced by a symmetry. It is
therefore not surprising that this kind of cancellation {\em cannot} be
achieved once corrections from two or more loops are included \cite{8}.}
We are thus looking for a symmetry that can {\em enforce} eqs.(\ref{e7})
and (\ref{e10}) (as well as their generalizations to gauge interactions).
In particular, we need equal numbers of physical (propagating) bosonic
and fermionic degrees of freedom, eq.(\ref{e7a}). In addition, we need
relations between various terms in the Lagrangian involving different
combinations of bosonic and fermionic fields, eqs.(\ref{e7b}) and
(\ref{e10}).

It is quite clear from these considerations that the symmetry we are looking
for must connect bosons and fermions. In other words, the generators $Q$
of this symmetry must turn a bosonic state into a fermionic one, and vice
versa. This in turn implies that the generators themselves carry half--integer
spin, i.e. are fermionic. This is to be contrasted with the generators
of the Lorentz group, or with gauge group generators, all of which are
bosonic. In order to emphasize the new quality of this new symmetry, which
mixes bosons and fermions, it is called {\it supersymmetry} (SUSY).

The simplest choice of SUSY generators is a 2--component (Weyl) spinor
$Q$ and its conjugate $\overline{Q}$. Since these generators are fermionic,
their algebra can most easily be written in terms of anti--commutators:
\ben \label{e12} \beq
\left\{ Q_\alpha , Q_\beta \right\} &= \left\{ \overline{Q}_{\dot{\alpha}} ,
\overline{Q}_{\dot{\beta}} \right\} = 0; \label{e12a} \\
\left\{ Q_\alpha ,  \overline{Q}_{\dot{\beta}} \right\} &= 2
\sigma_{\alpha \dot{\beta}}^{\mu} P_{\mu}; \ \ \ \ \ \ \ \
\left[Q_{\alpha}, P_{\mu} \right] = 0. \label{e12b}
\eeq \een
Here the indices $\alpha, \ \beta$ of $Q$ and $\dot{\alpha}, \
\dot{\beta}$ of $\overline{Q}$ take values 1 or 2, $\sigma^{\mu} = ( {\bf 1},
\sigma_i)$ with $\sigma_i$ being the Pauli matrices, and $P_{\mu}$ is 
the translation generator (momentum); it must appear in eq.(\ref{e12b}) for 
the SUSY algebra to be consistent with Lorentz covariance \cite{9}.

For a compact description of SUSY transformations, it will prove convenient
to introduce ``fermionic coordinates'' $\theta, \thb$. These are 
anti--commuting, ``Grassmann'' variables:
\be \label{e13}
\left\{ \theta, \theta \right\} = \left\{ \theta, \thb \right\} =
\left\{ \thb, \thb \right\} = 0.
\ee
A ``finite'' SUSY transformation can then be written as 
$\exp \left[ i ( \theta Q + \overline{Q} \thb - x_{\mu} P^{\mu} )
\right]$; this is to be compared with a non--abelian gauge transformation
$\exp \left( i \varphi_a T^a \right)$, with $T^a$ being the group generators.
Of course, the objects on which these SUSY transformations act must then
also depend on $\theta$ and \thb. This leads to the introduction of
{\em superfields}, which can be understood to be functions of $\theta$ and
\thb\ as well as the spacetime coordinates $x_{\mu}$. Since $\theta$ and
\thb\ are also two--component spinors, one can even argue that supersymmetry
doubles the dimension of spacetime, the new dimensions being fermionic.

For most purposes it is sufficient to consider infinitesimal SUSY 
transformations. These can be written as 
\be \label{e14}
\delta_S (\alpha, \alb) \Phi(x, \theta, \thb) = \left[ \alpha \frac
{\partial} {\partial \theta} + \alb \frac {\partial} {\partial \thb}
- i \left( \alpha \sigma_{\mu} \thb - \theta \sigma_{\mu} \alb
\right) \frac {\partial} {\partial x_{\mu}} \right] \Phi(x, \theta,\thb),
\ee
where $\Phi$ is a superfield and $\alpha, \ \alb$ are again Grassmann
variables. This corresponds to the following explicit representation of
the SUSY generators:
\be \label{e15}
Q_{\alpha} = \frac {\partial} {\partial \theta^{\alpha}} - i
\sigma^{\mu}_{\alpha \dot{\beta}} \thb^{\dot{\beta}} \partial_{\mu}; \ \ \ \
\overline{Q}_{\dot{\alpha}} = - \frac {\partial} {\partial \thb^{\dot{\alpha}}}
+ i \theta^{\beta} \sigma^{\mu}_{\beta \dot{\alpha}} \partial_{\mu}.
\ee
It will prove convenient to introduce SUSY--covariant derivatives, which
anti--commute with the SUSY transformation (\ref{e14}):
\be \label{e16}
D_{\alpha} = \frac {\partial} {\partial \theta^{\alpha}} + i
\sigma^{\mu}_{\alpha \dot{\beta}} \thb^{\dot{\beta}} \partial_{\mu}; \ \ \ \
\overline{D}_{\dot{\alpha}} = - \frac {\partial} {\partial \thb^{\dot{\alpha}}}
- i \theta^{\beta} \sigma^{\mu}_{\beta \dot{\alpha}} \partial_{\mu}.
\ee
Note that eqs.(\ref{e14})--({\ref{e16}) imply that $\alpha$ and $\theta$
have mass dimension $-1/2$, while $Q$ and $D$ have mass dimension $+1/2$.

Eqs.(\ref{e14})--(\ref{e16}) have been written in a form that treats $\theta$
and \thb\ on equal footing. It is often more convenient to use ``chiral''
representations, where $\theta$ and \thb\ are treated slightly differently
(I am suppressing spinor indices from now on):
\ben \label{e17} \beq
{}& \hspace*{1 cm} {\rm L-representation}: \nonumber \\
\delta_S \Phi_L &= \left( \alpha \frac {\partial} {\partial \theta} +
\alb \frac {\partial} {\partial \thb} + 2 i \theta \sigma^{\mu} \alb 
\partial_{\mu} \right) \Phi_L; \nonumber \\
D_L &= \frac {\partial} {\partial \theta} + 2 i \sigma^{\mu} \thb 
\partial_{\mu}; \ \ \ \overline{D}_L = - \frac {\partial} {\partial
\thb}. \label{e17a} \\
{}& \hspace*{1 cm} {\rm R-representation}: \nonumber \\
\delta_S \Phi_R &= \left( \alpha \frac {\partial} {\partial \theta} +
\alb \frac {\partial} {\partial \thb} - 2 i \alpha \sigma^{\mu} \thb 
\partial_{\mu} \right) \Phi_R; \nonumber \\
\overline{D}_R &= -\frac {\partial} {\partial \thb} - 2 i \theta \sigma^{\mu}  
\partial_{\mu}; \ \ \ D_R = \frac {\partial} {\partial \theta}. \label{e17b}
\eeq \een
Clearly, $\overline{D} \ (D)$ has a particularly simple form in the L (R)
representation. The following identity allows to switch between 
representations:
\be \label{e18}
\Phi(x, \theta, \thb) = \Phi_L( x_{\mu} + i \theta \sigma_{\mu} \thb,
\theta, \thb) = \Phi_R( x_{\mu} - i \theta \sigma_{\mu} \thb,
\theta, \thb).
\ee
So far everything has been written for arbitrary superfields $\Phi$.
However, we will only need two kinds of special superfields, which are
{\em irreducible} representations of the SUSY algebra; they will be
discussed in the following two subsections.

\subsection*{3b. Chiral Superfields}
The first kind of superfield we will need are {\it chiral superfields}.
This name is derived from the fact that the SM fermions are chiral, that
is, their left-- and right--handed components transfer differently under
\sym. We therefore need superfields with only two physical fermionic
degrees of freedom, which can then describe the left-- or right--handed
component of an SM fermion. Of course, the same superfields will also
contain bosonic partners, the {\it sfermions}.

The simplest way to construct such superfields is to require either
\ben \label{e19} \beq
\overline{D} \Phi_L &\equiv 0 \hspace*{2cm} (\Phi_L \ {\rm is \ 
left-chiral}) \hspace*{2cm} {\rm or} \label{e19a} \\
D \Phi_R &\equiv 0 \hspace*{2cm} (\Phi_R \ {\rm is \ right-chiral}).
\label{e19b} \eeq \een
Clearly these conditions are most easily implemented using the chiral
representations of the SUSY generators and SUSY--covariant
derivatives.  For example, eq.(\ref{e17a}) shows that in the
L--representation, eq.(\ref{e19a}) simply implies that $\Phi_L$ is
independent of \thb, i.e. $\Phi_L$ only depends on $x$ and
$\theta$. Recalling that $\theta$ is an anti--commuting Grassmann
variable, eq.(\ref{e13}), we can then expand $\Phi_L$ as:
\be \label{e20} 
\Phi_L(x, \theta) = \phi(x) + \sqrt{2} \theta^\alpha \psi_\alpha(x) 
+ \theta^\alpha \theta^\beta \epsilon_{\alpha \beta} F(x), 
\ee 
where summation over identical upper and lower indices is understood,
and $\epsilon_{\alpha \beta}$ is the anti--symmetric tensor in two
dimensions. Recall that $\theta$ has mass dimension $-1/2$. Assigning
the usual mass dimension $+1$ to the scalar field $\phi$ then gives
the usual mass dimension $+3/2$ for the fermionic field $\psi$, and
the unusual mass dimension $+2$ for the scalar field $F$; the
superfield $\Phi$ itself has mass dimension $+1$. The expansion
(\ref{e20}) is exact, since $\theta$ only has two components, and
eq.(\ref{e13}) implies that the square of any one component vanishes;
hence there cannot be any terms with three or more factors of
$\theta$. The fields $\phi$ and $F$ are complex scalars, while $\psi$
is a Weyl spinor. At first glance, $\Phi_L$ seems to contain four
bosonic degrees of freedom and only two fermionic ones; however, we
will see later on that not all bosonic fields represent physical
(propagating) degrees of freedom. The expression for $\Phi_R$ in the
R--representation is very similar; one merely has to replace $\theta$
by \thb.\footnote{Note that one can also write a left--chiral
superfield using the right--chiral representation of the SUSY
generators, and vice versa. The physical content of the fields remains
the same, of course, but the expressions become quite a bit more
lengthy. Nevertheless we will be forced to do this on one later
occasion.}

Applying the explicit form (\ref{e17a}) of the SUSY transformation to
the left--chiral superfield (\ref{e20}) gives:
\begin{eqnarray} \label{e21}
\delta_S \Phi_L &=& \sqrt{2} \alpha^\alpha \psi_\alpha + 2 \alpha^\alpha
\theta^\beta \epsilon_{\alpha \beta} F + 2 i \theta^\alpha
\sigma^\mu_{\alpha \dot{\beta}} \alb^{\dot{\beta}} \partial_\mu \phi
+ 2 \sqrt{2} i \theta^\alpha \sigma^\mu_{\alpha \dot{\beta}} 
\alb^{\dot{\beta}} \theta^\beta \partial_\mu \psi_\beta 
\nonumber \\
&\equiv& \delta_S \phi + \sqrt{2} \theta \delta_S \psi + \theta \theta
\delta_S F.
\end{eqnarray}
The first two terms of the first line of eq.(\ref{e21}) come from the
application of the $\partial / \partial \theta$ part of $\delta_S$, while
the last two terms come from the $\partial_\mu$ part; note that the
$\partial_\mu$ part applied to the last term in eq.(\ref{e20}) vanishes,
since it contains three factors of $\theta$. The second line of
eq.(\ref{e21}) is just the statement that the SUSY algebra should close,
i.e. a SUSY transformation applied to a left--chiral superfield
should again give a left--chiral superfield. It is easy to see that this
is true, since the first line of eq.(\ref{e21}) does not contain any
terms $\propto \thb$, so an expansion as in eq.(\ref{e20}) must be
applicable to it. Explicitly, we have:
\ben \label{e22} \beq
\delta_S \phi &= \sqrt{2} \alpha \psi 
\hspace*{4cm} ({\rm boson \rightarrow fermion}) \label{e22a} \\
\delta_S \psi &= \sqrt{2} \alpha F + i \sqrt{2} \sigma^\mu \alb \partial_\mu
\phi \hspace*{1.4cm} ({\rm fermion \rightarrow boson}) \label{e22b} \\
\delta_S F &= -i \sqrt{2} \partial_\mu \psi \sigma^\mu \alb
\hspace*{2.5cm} (F \rightarrow {\rm total \ derivative}) \label{e22c}
\eeq \een
Notice in particular the result (\ref{e22c}); it
implies that $\int d^4x F(x)$ is {\em invariant} under SUSY transformations,
assuming as usual that boundary terms vanish. We will come back to this point
in Sec.~3d.

\subsection*{3c. Vector Superfields}
The chiral superfields introduced in the previous subsection can
describe spin--0 bosons and spin--1/2 fermions, e.g. the Higgs boson
and the quarks and leptons of the SM. However, we also have to describe
the spin--1 gauge bosons of the SM. To this end one introduces
{\it vector superfields} $V$. They are constrained to be
self--conjugate:
\be \label{e23}
V(x, \theta, \thb) \equiv V^\dagger (x, \theta, \thb).
\ee
This leads to the following representation of $V$ in component form:
\begin{eqnarray} \label{e24}
V(x, \theta, \thb) &=& \left( 1 + \frac{1}{4} \theta \theta \thb \thb
\partial_\mu \partial^\mu \right) C(x) + \left( i \theta + \frac{1}{2}
\theta \theta \sigma^\mu \thb \partial_\mu \right) \chi(x)
+ \frac {i}{2} \theta \theta \left[ M(x) + i N(x) \right]
\nonumber \\
&+& \left( -i \thb + \frac {1}{2} \thb \thb \sigma^\mu \theta \partial_\mu
\right) \overline{\chi}(x) - \frac{i}{2} \thb \thb \left[ M(x) - i N(x)
\right] \nonumber \\
&-& \theta \sigma_\mu \thb A^\mu(x) + i \theta \theta \thb 
\overline{\lambda} (x) - i \thb \thb \theta \lambda(x) + \frac{1}{2}
\theta \theta \thb \thb D(x).
\end{eqnarray}
Here, $C, \ M, \ N$ and $D$ are real scalars, $\chi$ and $\lambda$ are
Weyl spinors, and $A^\mu$ is a vector field. If $A^\mu$ is to describe a 
gauge boson, $V$ must transform as an adjoint representation of the gauge
group.

The general form (\ref{e24}) is rather unwieldy. Fortunately, we now
have many more gauge degrees of freedom than in nonsupersymmetric
theories, since now the gauge parameters are themselves superfields.
A general non--abelian supersymmetric gauge transformation acting on $V$ can be
described by
\be \label{e25}
e^{g V} \longrightarrow e^{-ig \Lambda^\dagger} e^{gV}
e^{i g \Lambda}
\ee
where $\Lambda(x,\theta,\thb)$ is a chiral superfield and $g$ is the
gauge coupling. In the case of an abelian gauge symmetry, this
transformation rule can be written more simply as
\be \label{e26}
V \longrightarrow V + i (\Lambda - \Lambda^\dagger) \hspace*{2cm}
({\rm abelian \ case}).
\ee
Remembering that a chiral superfield contains four scalar (bosonic) degrees
of freedom as well as one Weyl spinor, it is quite easy to see that one can
use the transformation (\ref{e25}) or (\ref{e26}) to chose
\be  \label{e27}
\chi(x) = C(x) = M(x) = N(x) \equiv 0.
\ee
This is called the ``Wess--Zumino'' (W--Z) gauge; it is in some sense
the SUSY analog of the unitary gauge in ``ordinary'' field theory,
since it removes many unphysical degrees of freedom. Notice, however,
that we have only used three of the four bosonic degrees of freedom in
$\Lambda$. We therefore still have the ``ordinary'' gauge freedom,
e.g. according to $A_\mu(x) \longrightarrow A_\mu(x) + \partial_\mu
\varphi (x)$ for an abelian theory. In other words, the W--Z gauge can
be used in combination with any of the usual gauges. However, the
choice (\ref{e27}) is sufficient by itself to remove the first two
lines of eq.(\ref{e24}), leading to a much more compact expression for
$V$. Assigning the usual mass dimension $+1$ to $A^\mu$ gives the
canonical mass dimension $+3/2$ for the fermionic field $\lambda$,
while the field $D$ has the unusual mass dimension $+2$, just like the
$F-$component of the chiral superfield (\ref{e20}). Notice also that
the superfield $V$ itself has no mass dimension.

Applying a SUSY transformation to eq.(\ref{e24}) obviously gives a
lengthier expression than in case of chiral superfields. Here I only quote
the important result
\be \label{e28}
\delta_S D = - \alpha \sigma^\mu \partial_\mu \overline{\lambda}
+ \alb \sigma^\mu \partial_\mu \lambda,
\ee
which shows that the $D$ component of a vector superfield transforms into
a total derivative. Together with the analogous result (\ref{e22c}) for
chiral superfields, this provides the crucial clue for the construction
of the Lagrangian, to which we turn next.

\subsection*{3d. Construction of the Lagrangian}
We are now ready to attempt the construction of the Lagrangian of a
supersymmetric field theory. By definition, we want the action to be
invariant under SUSY transformations:
\be \label{e29}
\delta_S \int d^4x \lag (x) = 0.
\ee
This is satisfied if \lag\ itself transforms into a total derivative.
We saw in eqs.(\ref{e22a}) and (\ref{e28}) that the highest components
(those with the largest number of $\theta$ and \thb\ factors) of chiral
and vector superfields satisfy this requirement; they can therefore be
used to construct the Lagrangian. We can thus write the action $S$
schematically as
\be \label{e30}
S = \int d^4 x \left( \int d^2 \theta \lag_F + \int d^2 \theta d^2 \thb
\lag_D \right),
\ee
where the integration over Grassmann variables is defined as:
\be \label{e31}
\int d \theta_\alpha = 0, \hspace*{1.8cm}
\int \theta_\alpha d \theta_\alpha = 1
\ee
(no summation over $\alpha$). $\lag_F$ and $\lag_D$ in eq.(\ref{e30}) are
general chiral and vector superfields, giving rise to ``F--terms'' and
``D--terms'', respectively.

In order to make this more explicit, let us compute the product of two
left--chiral superfields:
\begin{eqnarray} \label{e32}
\Phi_{1,L} \Phi_{2,L} &=& \left( \phi_1 + \sqrt{2} \theta \psi_1
+ \theta \theta F_1 \right) \left( \phi_2 + \sqrt{2} \theta \psi_2
+ \theta \theta F_2 \right) \nonumber \\
&=& \phi_1 \phi_2 + \sqrt{2} \theta \left( \psi_1 \phi_2 +
\phi_1 \psi_2 \right) + \theta \theta \left( \phi_1 F_2 + \phi_2 F_1
- \psi_1 \psi_2 \right).
\end{eqnarray}
(Recall that $\theta \theta \theta = 0$.) Notice that this is itself a 
left--chiral superfield, since it does not depend on \thb, so it is a
candidate for a contribution to the $\lag_F$ term in the action (\ref{e30}).
Indeed, the very last term in eq.(\ref{e32}) looks like a fermion
mass term! We have thus identified a first possible contribution to the
Lagrangian.

Of course, if the product of two left--chiral superfields is a left--chiral
superfield, by induction the same must be true for the product of any number
of left--chiral superfields. Let us therefore compute the highest component
in the product of three such fields:
\be \label{e33}
\int d^2 \theta \Phi_{1,L} \Phi_{2,L} \Phi_{3,L} = \phi_1 \phi_2 F_3
+ \phi_1 F_2 \phi_3 + \phi_1 \phi_2 F_3 - \psi_1 \phi_2 \psi_3
- \phi_1 \psi_2 \psi_3 - \psi_1 \psi_2 \phi_3.
\ee
Note that the last three terms in eq.(\ref{e33}) describe Yukawa interactions
between one scalar and two fermions; in the SM such interactions give rise to
quark and lepton masses. We have thus identified our first interaction term
in the SUSY Lagrangian! Notice that if we, e.g., call $\phi_1$ the Higgs
field, and $\psi_2$ and $\psi_3$ the left-- and right--handed components
of the top quark, respectively\footnote{More exactly, $\psi_3$ describes
the left--handed anti--top.}, eq.(\ref{e33}) will not only produce the
desired Higgs--top--top interaction, but also interactions between a scalar
top $\tilde{t}$, the fermionic ``higgsino'' $\widetilde{h}$, and the
top quark, with {\em equal} strength. This is a first example of relations
between couplings enforced by supersymmetry.

So far we have identified terms that can give rise to explicit fermion
masses, eq.(\ref{e30}), as well as Yukawa interactions, eq.(\ref{e32}),
but we have not yet found any terms with derivatives, i.e. kinetic
energy terms. Clearly multiplying even more left--chiral superfields
with each other is not going to help; it is easy to see that this gives
rise to terms with mass dimension $>4$ in the Lagrangian, which lead to
non--renormalizable interactions. Let us instead consider the product
of a left--chiral superfield and its conjugate. The latter is in fact
a right--chiral superfield. Since we have to use the same representation
of the SUSY generators everywhere, we first have to write this right--chiral
superfield in the L--representation, using eq.(\ref{e18}):
\begin{eqnarray} \label{e34}
\left[ \Phi_L (x, \theta) \right]^\dagger &=& \phi^\ast - 2 i \theta
\sigma_\mu \thb \partial^\mu \phi^\ast - 2 \left( \theta \sigma_\mu \thb
\right)  \left( \theta \sigma_\nu \thb \right) \partial^\mu \partial^\nu
\phi^\ast \nonumber \\
&+& \sqrt{2} \thb \overline{\psi} - 2 \sqrt{2} i \left( \theta \sigma_\mu \thb
\right) \partial^\mu \left( \thb \overline{\psi} \right) + \thb \thb F^\ast.
\end{eqnarray}
Clearly the product $\Phi_L \Phi_L^\dagger$ is self--conjugate, i.e. it is
a vector superfield. It is therefore a candidate contribution to the
``D--terms'' in the action (\ref{e30}):
\be \label{e35}
\int d^2 \theta d^2 \thb \Phi_L \Phi_L^\dagger = F F^\ast - \phi 
\partial_\mu \partial^\mu \phi^\ast - i \overline{\psi} \sigma_\mu
\partial^\mu \psi.
\ee
This contains kinetic energy terms for the scalar component $\phi$
as well as the fermionic component $\psi$ of chiral superfields! 
Equally importantly, eq.(\ref{e35}) does {\em not} contain kinetic
energy terms for $F$. This field does therefore not propagate; it is
a mere auxiliary field, which can be integrated out exactly using its purely
algebraic equation of motion. A chiral superfield therefore only has two
{\em physical} bosonic degrees of freedom, described by the complex
scalar $\phi$, i.e. it contains equal numbers of propagating bosonic
and fermionic degrees of freedom.

In order to illustrate how the $F-$fields can be removed from the
Lagrangian, let us introduce the {\it superpotential} $f$:
\be \label{e36}
f(\Phi_i) = \sum_i k_i \Phi_i + \frac{1}{2} \sum_{i,j} m_{ij}
\Phi_i \Phi_j + \frac{1}{3} \sum_{i,j,k} g_{ijk} \Phi_i \Phi_j \Phi_k,
\ee
where the $\Phi_i$ are all left--chiral superfields, and the
$k_i, \ m_{ij}$ and $g_{ijk}$ are constants with mass dimension
2, 1 and 0, respectively. The contributions to the Lagrangian that we
have identified so far can be written compactly as
\begin{eqnarray} \label{e37}
\lag &=& \sum_i \int d^2 \theta d^2 \thb \Phi_i \Phi_i^\dagger
+ \left[ \int d^2 \theta f(\Phi_i) + h.c. \right]
\nonumber \\
&=& \sum_i \left( F_i F_i^\ast + \left| \partial_\mu \phi \right|^2
- i \overline{\psi}_i \sigma_\mu \partial^\mu \psi_i \right)
\nonumber \\
&+& \left[ \sum_j \frac {\partial f(\phi_i)} {\partial \phi_j}
F_j - \frac{1}{2} \sum_{j,k} \frac {\partial^2 f(\phi_i)} {\partial \phi_j
\partial \phi_k} \psi_j \psi_k + h.c. \right].
\end{eqnarray}
Note that in the last line of eq.(\ref{e37}), $f$ is understood to be a
function of the scalar fields $\phi_i$, rather than of the superfields
$\Phi_i$. Using eq.(\ref{e36}) it is easy to convince oneself that
the last line in eq.(\ref{e37}) indeed reproduces the previous results
(\ref{e32}) and (\ref{e33}). Let us now integrate out the auxiliary
fields $F_j$. Their equations of motion are simply given by
$\partial \lag / \partial F_j = 0$, which implies
\be \label{e38}
F_j = - \left[ \frac {\partial f(\phi_i)} {\partial \phi_j} \right]^\ast.
\ee
Plugging this back into eq.(\ref{e37}) gives:
\be \label{e39}
\lag = \lag_{\rm kin} - \left[ \sum_{j,k} \frac {\partial^2 f(\phi_i)}
{\partial \phi_j \partial \phi_k} \psi_j \psi_k + h.c. \right] - \sum_j
\left| \frac {\partial f(\phi_i)} {\partial \phi_j} \right|^2,
\ee
where $\lag_{\rm kin}$ stands for the second line in eq.(\ref{e37}).
The second term in the Lagrangian (\ref{e39}) describes fermion masses
and Yukawa interactions, while the last term describes scalar mass
terms and scalar interactions. Since both terms are determined by the
single function $f$, there are clearly many relations between coupling
constants.

Before elaborating on this last point, we introduce gauge interactions.
The coupling of the gauge (super)fields to the (chiral) matter (super)fields
is accomplished by a SUSY version of the familiar ``minimal coupling'':
\begin{eqnarray} \label{e40}
\int d^2 \theta d^2 \thb \Phi^\dagger \Phi & \longrightarrow & \int
d^2 \theta d^2 \thb \Phi^\dagger e^{2 g V} \Phi \nonumber \\
&=& \left| D_\mu \phi \right|^2 - i \overline{\psi} \sigma_\mu
D^\mu \psi + g \phi^\ast D \phi + i g \sqrt{2} \left( \phi^\ast \lambda \psi
- \overline{\lambda} \overline{\psi} \phi \right) + |F|^2.
\end{eqnarray}
In the second step I have used the W--Z gauge (\ref{e27}), and
introduced the usual gauge--covariant derivative $D_\mu = \partial_\mu
+ i g A_\mu^a T_a$, where the $T_a$ are group generators. Note that
this piece of the Lagrangian not only describes the interactions of
the matter fields (both fermions and scalars) with the gauge fields,
but also contains gauge--strength Yukawa--interactions between
fermions (or higgsinos) $\psi$, sfermions (or Higgs bosons) $\phi$, and 
gauginos $\lambda$.

Finally, the kinetic energy terms of the gauge fields can be described with
the help of the superfield
\be \label{e41}
W_\alpha = \left( \overline{D}_{\dot{\alpha}} \overline{D}_{\dot{\beta}}
\epsilon^{\dot{\alpha} \dot{\beta}} \right) e^{-gV} D_\alpha e^{gV};
\ee
the $D, \overline{D}$ appearing here are again SUSY--covariant derivatives,
which carry spinor subscripts. For abelian symmetries, this reduces to
$W_\alpha = \left( \overline{D}_{\dot{\alpha}} \overline{D}_{\dot{\beta}}
\epsilon^{\dot{\alpha} \dot{\beta}} \right) D_\alpha V$. Since
$\overline{D}_{\dot{\alpha}} \overline{D}_{\dot{\alpha}} \equiv 0, \ 
\overline{D}_{\dot{\alpha}} W_\alpha = 0$, so $W_\alpha$ is a left--chiral
superfield; its behaviour under gauge transformations is identical to
that of $e^{gV}$, see eq.(\ref{e25}). One can show that the product
$W_\alpha W^\alpha$ is gauge invariant; as shown earlier, it is also a
left--chiral superfield, so its $\theta \theta$ component may appear in
the Lagrangian:
\begin{eqnarray} \label{e42}
\frac{1}{32 g^2} W_\alpha W^\alpha &=& - \frac{1}{4} F^a_{\mu\nu}
F_a^{\mu\nu} + \frac {1}{2} D_a D^a \nonumber \\
&+& \left( - \frac {i}{2} \lambda^a \sigma_\mu \partial^\mu 
\overline{\lambda}_a + \frac {1}{2} g f^{abc} \lambda_a \sigma_\mu
A^\mu_b \overline{\lambda}_c + h.c. \right).
\end{eqnarray}
In addition to the familiar kinetic energy term for the gauge fields,
this also contains a kinetic energy terms for the gauginos $\lambda_a$, as
well as the canonical coupling of the gauginos to the gauge fields, which is
determined by the group structure constants $f^{abc}$.

Note that eq.(\ref{e42}) does not contain a kinetic energy term for the
$D_a$ fields. They are therefore also auxiliary fields, and can again easily
be integrated out. From eqs.(\ref{e40}) and (\ref{e42}) we see that their
equation of motion is
\be \label{e43}
D_a = - g \sum_{i,j} \phi_i^\ast T_a^{ij} \phi_j,
\ee
where the group indices have been written explicitly. (The field $D$ in
eq.(\ref{e40}) is equal to $\sum_a D_a T^a$, in complete analogy to the
gauge fields.) The third term in the second line of eq.(\ref{e40}) and
the second term in eq.(\ref{e42}) then combine to give a contribution
\be \label{e44}
- V_D = - \frac {1}{2} \sum_a \left| \sum_{i,j} g \phi_i^\ast T^a_{ij}
\phi_j \right|^2
\ee
to the scalar interactions in the Lagrangian; these interactions are
completely fixed by the gauge couplings. This completes the construction
of the Lagrangian for a renormalizable supersymmetric field theory.

\subsection*{3e. Quadratic Divergencies, Part 2}
As a first application of the results of the previous subsection, let us
check that there are indeed no quadratic divergencies, at least at
one--loop order. It is easy to see that there are no quadratic divergencies 
from Yukawa interactions, since eqs.(\ref{e7}) hold. Eq.(\ref{e7a}) is
satisfied because, as emphasized in the paragraph below eq.(\ref{e35}),
each chiral superfield contains equal numbers of physical bosonic and fermionic
degrees of freedom. Eq.(\ref{e7b}) can be checked by inserting the relevant
part of the superportential $f_{\rm top} = \lambda_t T_L T_R H$,
where $T_L$ and $T_R$ contain the left-- and right--handed components of
the top quark as well as the corresponding squarks, into eq.(\ref{e39}).
In fact, eq.(\ref{e39}) even satisfies the more stringent requirements
(\ref{e10}); the Yukawa contribution to $\piff(0)$ therefore vanishes
identically.\footnote{The alert reader may have noticed that the contribution
(\ref{e44}) to the gauge interactions might produce contributions to
$m_{\tilde f}^2$ that are proportional to $m^2_W$ or $m_Z^2$, thereby
violating the condition (\ref{e10a}). We will come back to this point
shortly.}

We can now also compute the contribution from gauge interactions to the
Higgs two--point function, using eqs.(\ref{e40}), (\ref{e42}) and
(\ref{e44}). I will for the moment stick to the assumption that there is
only one Higgs doublet; this will later prove to be not entirely realistic,
but it is sufficient for the time being. As a further simplification,
I will ``switch off'' hypercharge interactions, so that $m_W = m_Z$.
Finally, I will use Feynman gauge, so contributions from the unphysical
would--be Goldstone bosons have to be included; their mass is equal
to $m_W$ in this gauge.

\begin{center}
\begin{picture}(400,100)(0,0)
\DashLine(20,50)(180,50){4}
\PhotonArc(100,78)(25,0,360){4}{14}
\DashLine(220,50)(270,50){4}
\DashLine(320,50)(370,50){4}
\PhotonArc(295,50)(25,0,360){4}{14}
\Text(10,50)[]{$\phi$}
\Text(65,75)[]{$V$}
\Text(210,50)[]{$\phi$}
\Text(380,50)[]{$\phi$}
\Text(295,15)[]{$V$}
\Text(295,85)[]{$V$}
\end{picture}

{\bf Fig.~5:} Gauge boson loop contributions to the Higgs self energy. $V$
stands for either $W^\pm$ or $Z$.
\end{center}

There are two types of contribution that involve only gauge bosons $V$ in
the loop, see Fig.~5:
\be \label{e45}
\piff^V(0) = N(V) \int \ddk \frac {-i g_{\mu\nu} } {k^2 -m_W^2} i \frac {g^2}
{2} g^{\mu\nu} = 3 g^2 \int \ddk \frac {1} {k^2-m_W^2}.
\ee
\begin{eqnarray} \label{e46}
\piff^{VV}(0) &=& N(V) \int \ddk \left( i g_{\mu\nu} g m_W \right) 
\left( i g_{\rho\sigma} g m_W \right) \frac { \left( -i g^{\mu\rho}
\right) \left( - i g^{\nu\sigma} \right) } { \left( k^2 - m_W^2 \right)^2}
\nonumber \\
&=& 6 g^2 m_W \int \ddk \frac {1} {\left( k^2 - m_W^2 \right)^2}.
\end{eqnarray}
Here and in the subsequent expressions, the superscripts of $\piff$
denote the particle(s) in the loop. Further, the effective number of
vector bosons $N(V) = 3/2$, since $Z$ boson loops get a suppression
factor 1/2 for identical particles.

\begin{center}
\begin{picture}(300,100)(0,0)
\DashLine(75,50)(125,50){4}
\DashLine(175,50)(225,50){4}
\PhotonArc(150,50)(25,0,180){4}{8.5}
\DashCArc(150,50)(25,180,360){4}
\Text(150,15)[]{$G$}
\Text(150,85)[]{$V$}
\Text(65,50)[]{$\phi$}
\Text(235,50)[]{$\phi$}
\end{picture}

{\bf Fig.~6:} Contributions to the Higgs self energy from loops involving a
gauge bosons $V=W^\pm$ or $Z$ and a would--be Goldstone boson $G=G^\mp$ or
$G^0$.
\end{center}

There are also contributions with a gauge boson and a would--be
Goldstone boson $G$ in the loop, as shown in Fig.~6:
\ben \label{e47} \beq
\piff^{Z G^0} (0) &= - \frac {g^2}{4} \int \ddk \frac {k^2}
{\left( k^2 - m_W^2 \right)^2}; \label{e47a} \\
\piff^{W^\pm G^\mp}(0) &= - \frac {g^2}{2} \int \ddk \frac {k^2}
{\left( k^2 - m_W^2 \right)^2}. \label{e47b}
\eeq \een
Note that eq.(\ref{e47b}) contains a factor of 2, since $W^+ G^-$ and
$W^- G^+$ loops are distinct.

\begin{center}
\begin{picture}(400,100)(0,0)
\DashLine(20,50)(180,50){4}
\DashCArc(100,75)(25,0,360){4}
\DashLine(220,50)(270,50){4}
\DashLine(320,50)(370,50){4}
\DashCArc(295,50)(25,0,360){4}
\Text(10,50)[]{$\phi$}
\Text(65,75)[]{$H$}
\Text(210,50)[]{$\phi$}
\Text(380,50)[]{$\phi$}
\Text(295,15)[]{$H$}
\Text(295,85)[]{$H$}
\end{picture}

{\bf Fig.~7:} Contributions to the Higgs self energy involving Higgs self
interactions. $H$ can be the physical Higgs field $\phi$ or one of the
would--be Goldstone modes $G^\pm$ or $G^0$.
\end{center}

The last bosonic contributions come from Higgs self--interactions, see
Fig.~7. It is important to note that in a supersymmetric model with
only Higgs doublets it is {\em impossible} to introduce Higgs
self--couplings through the superpotential $f$ of eq.(\ref{e36}). Such
an interaction would come from a cubic term in $f$, which is forbidden
by gauge invariance. The only Higgs self--interactions therefore come
from eq.(\ref{e44}). Focussing on the Higgs doublet field $H \equiv
\left( [ \phi + v + i G^0 ] / \sqrt{2}, G^- \right)$, this term reads:
\begin{eqnarray} \label{e48}
- V_D &=& - \frac {1}{8} g^2 \left[
\left( H_i^\ast \sigma_1^{ij} H_j \right)^2 + 
\left( H_i^\ast \sigma_2^{ij} H_j \right)^2 + 
\left( H_i^\ast \sigma_3^{ij} H_j \right)^2 \right] \nonumber \\
&=& - \frac {1}{8} g^2 \left[ \frac{1}{2} \left( \phi + v \right)^2
+ \frac{1}{2} \left( G^0 \right)^2 + \left| G^- \right|^2 \right]^2;
\end{eqnarray}
recall that the properly normalized $SU(2)$ generators are 1/2 times
the Pauli matrices. From eq.(\ref{e48}) one finds the following
contributions to the Higgs two--point function:
\ben \label{e49} \beq
\piff^\phi (0) &= \frac {3}{8} g^2 \int \ddk \frac{1} {k^2 - m^2_\phi};
\label{e49a} \\
\piff^{G^0} (0) &= \frac {1}{8} g^2 \int \ddk \frac {1} {k^2 - m_W^2};
\label{e49b} \\
\piff^{G^\pm} (0) &= \frac {1}{4} g^2 \int \ddk \frac {1} {k^2 - m_W^2}.
\label{e49c} 
\eeq \een
\ben \label{e50} \beq
\piff^{\phi\phi} (0) &= \frac {9}{8} g^2 m_W^2 \int \ddk \frac {1}
{ \left( k^2 - m_\phi^2 \right)^2} ; \label{e50a} \\
\piff^{G^0 G^0} (0) &= \frac {1}{8} g^2 m_W^2 \int \ddk \frac {1}
{ \left( k^2 - m_W^2 \right)^2} ; \label{e50b} \\
\piff^{G^+ G^-} (0) &= \frac {1}{4} g^2 m_W^2 \int \ddk \frac {1}
{ \left( k^2 - m_W^2 \right)^2}, \label{e50c}
\eeq \een
where I have used $g v = 2 m_W$; the contributions (\ref{e49a},b)
and (\ref{e50a},b) again contain factors 1/2 due to identical particle
loops.

Clearly the bosonic contributions give a nonvanishing quadratic
divergence, given by the sum of eqs.(\ref{e45}), (\ref{e47}) and
(\ref{e49}). However, there are also fermionic contributions involving
the higgsinos and gauginos. Their coupling to the Higgs field is
determined by the next--to--last term in the second line of
eq.(\ref{e40}), which couples the Higgs scalar to a gaugino and a
higgsino. In the presence of gauge symmetry breaking, $v \neq 0$, this
interaction also gives rise to a mass term between the gaugino and
higgsino fields. Recall that the spinors in eq.(\ref{e40}) are 
two-component Weyl spinors. Such an off--diagonal mass term between
Weyl spinors can be understood as a diagonal (Dirac) mass term for a
four--component spinor that contains the two Weyl--spinors as its
left-- and right--handed components. In the present case this produces
a charged ``chargino'' $\wt{W}$ with mass $\sqrt{2} m_W$, as well as
a (Dirac) ``neutralino'' $\wt{Z}$ with mass $m_W$. (Recall that hypercharge
interactions are switched off for the time being.) The chargino field
couples to the Higgs boson $\phi$ with strength $g/\sqrt{2}$, while
the neutralino couples with strength $g/2$. This relative factor
of $\sqrt{2}$ between the mass and coupling of the chargino and those
of the neutralino follows from the fact that the charged $SU(2)$
gauginos are given by $(\lambda_1 \pm i \lambda_2)/\sqrt{2}$, while
the neutral gaugino is simply $\lambda_3$. This gives the following
fermionic contributions, see eq.(\ref{e4}):
\ben \label{e51} \beq
\piff^{\wt{W} \wt{W}} (0) &= - 2 g^2 \int \ddk \left[
\frac {1} {k^2 - 2 m_W^2} + \frac {4 m_W^2} {\left( k^2 - 2 m_W^2
\right)^2} \right]; \label{e51a} \\
\piff^{\wt{Z} \wt{Z}} (0) &= - g^2 \int \ddk \left[
\frac {1} {k^2 - m_W^2} + \frac {2 m_W^2} {\left( k^2 - m_W^2
\right)^2} \right]. \label{e51b}
\eeq \een

We see that the total quadratic divergence does indeed cancel! The sum of
eqs.(\ref{e51}) cancels the contribution (\ref{e45}), while the contribution
from eqs.(\ref{e47}) cancels that from eqs.(\ref{e49}). The contributions
from eqs.(\ref{e46}) and (\ref{e50}) are by themselves only logarithmically
divergent. 

Before breaking out the champagne, we should check that we have not missed
any terms -- and, indeed, we have! The scalar self--interaction (\ref{e44})
also contains terms
\be \label{e52}
\lag_{\phi \tilde{f}} = \frac {g^2}{2} \left( \frac{1}{2} \phi^2 + v
\phi \right) \sum_f I_{3,f} \left| \tilde{f} \right|^2,
\ee
which leads to a total contribution from sfermion tadpole diagrams:
\be \label{e53}
\piff^{\tilde f} (0) = \frac {g^2}{2} \sum_f I_{3,f} \int \ddk
\frac {1} {k^2 - m^2_{\tilde f}}.
\ee
Here, $I_{3,f}$ is the weak isospin of fermion $f$ (which is identical to that
of its bosonic superpartner $\tilde{f}$, of course). Fortunately the trace
of $I_3$ over a complete representation of $SU(2)$ vanishes, so the contribution
(\ref{e53}) is in fact not quadratically divergent. Consider the case of a
single $SU(2)$ doublet. Let $m_+$ be the mass of of the $I_3 = +1/2$ 
sfermion. Gauge invariance implies that the $I_3 = -1/2$ sfermion must
have the same mass, except for the contributions from eq.(\ref{e44}) due to
the spontaneous breaking of $SU(2)$. The mass of the $I_3 = -1/2$ sfermion
is then in total given by $m_+^2 - m_W^2$, and the contribution to 
eq.(\ref{e53}) becomes:
\begin{eqnarray} \label{e54}
\left. \piff^{\tilde f} (0) \right|_{1 \ {\rm doublet}} &=& \frac {g^2}{4}
\int \ddk \left( \frac {1} {k^2 - m_+^2} - \frac{1} {k^2 - m_+^2 + m_W^2}
\right) \nonumber \\
&=& \frac {g^2}{4} \int \ddk \frac {m_W^2} { \left( k^2 - m_+^2 \right)
\left( k^2 - m_+^2 + m_W^2 \right) } ,
\end{eqnarray}
which is only logarithmically divergent.

So far, so good. We are clearly on the right track: There is no one--loop
quadratic divergence from $SU(2)$ interactions. This proof can be extended
to all orders using ``supergraphs'', i.e. Feynman rules for superfields
\cite{6}. However, disaster strikes when we try to re--introduce hypercharge
interactions. Much of the calculation presented here still goes through, with
minor changes. However, at the end a nonvanishing divergence remains, if
the model only contains one Higgs doublet and any number of complete 
(s)fermion generations of the SM. The problem can be traced back to the
fact that in this case the total trace of the hypercharge generator does
not vanish. Its trace over a complete (s)fermion generation does vanish,
but this leaves the contribution from the single Higgs doublet. Clearly
the model we have been using in this subsection is still not fully
realistic.\footnote{Hypercharge loop contributions to the Higgs mass have
another peculiarity. As already mentioned, the trace of the hypercharge
over a complete generation vanishes, given an only logarithmically divergent
$\tilde{f}$ contribution in analogy with eq.(\ref{e53}). However, at
least in most models the masses of different sfermions within the same
generation are not related by gauge invariance, so there could in general
be (very) large mass splittings. The scale of the log--divergent and
finite contributions to $\piff$ would then be set by these mass splittings,
not by $m_W^2$. This becomes a concern if one tries to push the masses of
the sfermions of the first two generations to very large values \cite{10,10a},
which otherwise need not lead to unacceptably large corrections to $\piff$,
due to the smallness of the first and second generation Yukawa couplings. 
Finally, in the context of supergravity or superstring theory, an anomalous
$U(1)$ factor can be part of a consistent and phenomenologically acceptable
model; see e.g. the first ref.\cite{10a}.}

Finally, notice that even in the absence of hypercharge interactions the
total gauge contribution to $\piff(0)$ does not vanish; in fact, some
logarithmic divergencies remain. This is related to the fact that in the
model considered here, one cannot break \sym\ invariance without breaking
supersymmetry. This brings us to the issue of supersymmetry breaking, to
which we turn next.

\subsection*{3f. Supersymmetry Breaking}
As emphasized earlier, the supersymmetric Lagrangian constructed in 
Sec.~3d satisfies eq.(\ref{e10a}), that is, the masses of the ``ordinary''
SM particles and their superpartners are identical. This is clearly not
realistic; there is no selectron with mass 511 keV, nor is there a smuon
with mass 106 MeV, etc. Indeed, as mentioned in the Introduction, no
superpartners have been discovered yet. Searches at the $e^+e^-$ collider
LEP imply that all charged sparticles must be heavier than 60 to 80 GeV
\cite{11}. Similarly, searches at the Tevatron $p \bar{p}$ collider
imply bounds on squark and gluino masses between 150 and 220 GeV \cite{12}.
Hence supersymmetry must be broken.

The great success of the Standard Model with its broken \sym\ symmetry
should have convinced everyone of the usefulness of broken symmetries.
Unfortunately it is not easy to break supersymmetry spontaneously. One
problem follows directly from the definition of the SUSY algebra, 
eq.(\ref{e12b}), which implies
\be \label{e55}
\frac{1}{4} \left( \qbar_1 Q_1 + Q_1 \qbar_1 + \qbar_2 Q_2 + Q_2 \qbar_2
\right) = P^0 \equiv H \geq 0,
\ee
where $H$ is the Hamiltonian (energy operator). The fact that this is
non--negative simply follows from it being a sum of perfect
squares. If the vacuum state $| 0 \rangle$ is supersymmetric, then
$Q_\alpha |0 \rangle = \qbar_{\dot{\alpha}} |0 \rangle = 0$, and
eq.(\ref{e54}) implies $E_{\rm vac} \equiv \langle 0 | H | 0 \rangle =
0$. On the other hand, if the vacuum state is {\em not}
supersymmetric, i.e. at least one SUSY generator does not annihilate
the vacuum, then eq.(\ref{e55}) implies $E_{\rm vac} > 0$. In other words,
global supersymmetry can only be broken spontaneously if there is a
positive vacuum energy. This might give rise to a troublesome cosmological
constant, although the connection between a microscopic vacuum energy and
a macroscopic cosmological constant is not entirely straightforward
\cite{13}.

As an example of the general result (\ref{e55}), consider the case where
supersymmetry is broken by the vev of some scalar particle, in direct analogy
to \sym\ breaking in the SM. The scalar potential contains two pieces,
given in eqs.(\ref{e39}) and (\ref{e44}):
\be \label{e56}
V = \sum_i \left| \frac {\partial f} {\partial \phi_i} \right|^2
 + \sum_l \frac {g_l^2}{2} \sum_a \left| \sum_{i,j} \phi_i^\ast
T^{ij}_{l,a} \phi_j \right|^2,
\ee
where $l$ labels the simple groups whose product forms the entire gauge
group of the model (e.g., $SU(3) \times \sym$ in the SM). We see that
indeed $V \geq 0$. We can therefore break SUSY if either $\langle 
F_i \rangle = \langle \partial f / \partial \phi_i \rangle \neq 0$ for
some $i$ [``F--term breaking'', see eq.(\ref{e38})], or if
$\langle D_{l,a} \rangle = \langle \sum_{i,j}  \phi_i^\ast
T^{ij}_{l,a} \phi_j \rangle \neq 0$ for some combination $(l,a)$
[(``D--term breaking'', see eq.(\ref{e43})]; in the latter case some
gauge symmetries will be broken as well. In fact, the example we
discussed in the previous subsection has D--term breaking, since
the D--term associated with the $I_3$ generator has a nonvanishing
vev, see eq.(\ref{e48}); this explains why the total contribution
to $\piff$ did no vanish in this example. However, clearly the second
term in eq.(\ref{e56}) can be minimized (set to zero) if all vevs vanish,
$\langle \phi_i \rangle = 0$ for all $i$. Turning the symmetry breaking
point into the absolute minimum of the potential therefore requires
nontrivial contributions from the first term in eq.(\ref{e56}).

The construction of realistic models with spontaneously broken SUSY is
made even more difficult by the fact that in such models eq.(\ref{e10a})
still remains satisfied ``on average''. More exactly, the supertrace over
the whole mass matrix vanishes in more with pure F--term breaking \cite{14}:
\be \label{e57}
{\rm Str} {\cal M}^2 \equiv \sum_J (-1)^{2J} {\rm tr} {\cal M}^2_J = 0,
\ee
where $J$ is the spin, and ${\cal M}_J$ is the mass matrix for all
particles with spin $J$. This is problematic, because we want {\em
all} sfermions to be significantly heavier than their SM partners
(with the possible exception of the scalar top). In principle one
could still satisfy the constraint (\ref{e57}) by making the gauginos
quite heavy; unfortunately this seems almost impossible to achieve in
practice. All potentially realistic globally supersymmetric models of
spontaneous SUSY breaking where sparticles get masses at tree--level
therefore contain a new $U(1)$ whose D--term is nonzero in the minimum
of the potential, as well as a rather large number of superfields
beyond those required by the field content of the SM \cite{6a}.
Recent models that attempt to break global SUSY spontaneously instead
circumvent the constraint (\ref{e57}) by creating most sparticle
masses only through radiative corrections \cite{15}; this also
necessitates the introduction of several additional superfields.

Most phenomenological analyses therefore do not attempt to understand SUSY
breaking dynamically; rather, it is parametrized by simply inserting
``soft breaking terms'' into the Lagrangian. ``Soft'' here means that
we want to maintain the cancellation of quadratic divergencies; e.g. we
want to respect eqs.(\ref{e7}). The explicit calculation of Sec.~2 showed
that, at least to one--loop order, quadratic divergencies still cancel
even if we introduce
\begin{itemize}
\item scalar mass terms $- m^2_{\phi_i} \left| \phi_i \right|^2$, and
\item trilinear scalar interactions $- A_{ijk} \phi_i \phi_j \phi_k 
+ h.c.$
\end{itemize}
into the Lagrangian. Girardello and Grisaru \cite{16} have shown that this
result survives in all orders in perturbation theory. They also identified
three additional types of soft breaking terms:
\begin{itemize}
\item gaugino mass terms $- \frac{1}{2} m_l \bar{\lambda}_l \lambda_l$,
where $l$ again labels the group factor;
\item bilinear terms $- B_{ij} \phi_i \phi_j + h.c.$; and
\item linear terms $-C_i \phi_i$.
\end{itemize}
Of course, linear terms are gauge invariant only for gauge singlet 
fields.\footnote{Under certain circumstances one can also introduce
trilinear interactions of the form $\tilde{A}_{ijk} \phi_i \phi_j
\phi_k^\ast + h.c.$ \cite{17}.} Note that we are {\em not} allowed
to introduce additional masses for chiral fermions, beyond those
contained in the superpotential. Also, the relations between dimensionless
couplings imposed by supersymmetry must not be broken.

This completes our discussion of the construction of ``realistic''
supersymmetric field theories. Let us now apply these results to the
simplest such model.

\section*{4. The Minimal Supersymmetric Standard Model}
Let us now try to construct a fully realistic SUSY model, i.e. a theory with
softly broken supersymmetry that satisfies all phenomenological constraints.
As already emphasized repeatedly, the main motivation for introducing
weak--scale supersymmetry is the absence of quadratic divergencies, which
leads to a solution of the (technical aspect of the) hierarchy problem.
There are, however, further arguments why supersymmetric theories might be
interesting. One is based on the Haag--Lopuszanski--Sohnius (HLS) theorem
\cite{9}; it states that the biggest symmetry which an interacting, unitary
field theory can have is the direct product of a (possibly very large) gauge
symmetry, Lorentz invariance, and (possibly extended) supersymmetry. The
first two ingredients are part of the highly succesful Standard Model; this
naturally raises the question whether making use of the third kind of
symmetry allowed by the HLS theorem leads to an even better description of
Nature.

Furthermore, supersymmetry appears very naturally in superstring theory. Often
the existence of space--time supersymmetry is even considered to be a firm 
prediction of string theory. String theory, in turn, is clearly our currently
best hope for a ``theory of everything'', which would, in particular, include
a quantum theory of gravity. However, this argument only requires supersymmetry
at or below the Planck scale, not necessarily at the weak scale.

These two arguments are admittedly rather speculative. A more
practical advantage of supersymmetric theories becomes apparent when
we compare them with their main competitor, technicolor models
\cite{18}. In these models one tries to solve the problem of quadratic
divergencies by dispensing with elementary scalars altogether. The
Higgs mechanism is then replaced by a non--perturbative mechanism,
where a confined ``technicolor'' gauge interaction leads to the
formation of ``techniquark'' condensates, which break (local) \sym\
invariance in a way reminiscent of the breaking of the (global) chiral
symmetry of QCD by quark condensates. I personally find the Higgs
mechanism much more elegant and innovative, but many of my colleagues
seem to consider the technicolor idea to be at least in principle more
appealing, since it appears to give a more dynamical understanding of
gauge symmetry breaking. In practice this hope is not really borne
out, however: Since gauge symmetry breaking is assumed to be due to
some non--perturbative dynamics, it is very difficult to make firm
predictions for physical observables. The lessons learned from the
study of low--energy hadron physics unfortunately turned out to be
rather useless here, since a successful technicolor theory must not be
a scaled--up version of QCD; such a theory would have much too strong
flavor changing neutral currents, and give too large contributions to
certain electroweak precision variables, most notably the
``S--parameter'' \cite{19}.

In contrast, supersymmetric theories might allow a perturbative description of
Nature at energies between about 1 GeV and the Planck scale. Furthermore, it
is quite easy to construct a potentially realistic SUSY model, as will be
demonstrated in the subsequent subsection.

\subsection*{4a. Definition of the Model}
As implied by the name, the minimal supersymmetric standard model
(MSSM) is essentially a straightforward supersymmetrization of the SM.
In particular, ``minimal'' means that we want to keep the number of
superfields and interactions as small as possible. Since the SM matter
fermions reside in different representations of the gauge group than
the gauge bosons, we have to place them in different superfields; no
SM fermion can be identified as a gaugino.\footnote{One occasionally
sees the statement that supersymmetry links gauge and matter fields.
This is not true in the MSSM, nor in any potentially realistic SUSY
model I know.} One generation of the SM is therefore described by five
left--chiral superfields: $Q$ contains the quark and squark $SU(2)$
doublets, $U^c$ and $D^c$ contains the (s)quark singlets, $L$ contains
the (s)lepton doublets, and $E^c$ contains the (s)lepton singlets.
Note that the $SU(2)$ singlet superfields contain left--handed
anti--fermions; their scalar members therefore have charge $+1$ for
\ter, $-2/3$ for \tur, and $+1/3$ for \tdr. Of course, we need three
generations to describe the matter content of the SM.

As discussed in Sec.~3, we have to introduce vector superfields to describe the
gauge sector. In particular, we need eight gluinos \tg\ as partners of the 
eight gluons of QCD, three winos $\wt{W}$ as partners of the $SU(2)$ gauge
bosons, and a bino $\wt{B}$ as $U(1)_Y$ gaugino. Since \sym\ is broken, the
winos and the bino are in general not mass eigenstates; rather, they mix with
fields with the same charge but different \sym\ quantum numbers. We will come
back to this point in Sec.~4c.

The only slight subtlety in the field content of the MSSM is in the
choice of the Higgs sector. As in the SM, we want to break \sym\
invariance by $SU(2)$ doublet scalars with hypercharge $|Y|=1/2$.
Looking through the fields that have already been introduced, one
notices that the slepton doublets \tll\ fulfill this requirement. It
is thus natural to ask whether the sneutrino fields can play the role
of the Higgs boson of the SM. Unfortunately the answer is No
\cite{20}. The terms required to give masses to the charged leptons
explicitly break lepton number, if the sneutrinos were to serve as
Higgs fields. This leads to a host of problems. The most stringent
bound on doublet sneutrino vevs comes from the requirement that all
neutrino masses must be very small \cite{21}, which implies $\langle
\tilde{\nu} \rangle^2 \ll M_Z^2$ \cite{22}.

We therefore have to introduce dedicated Higgs superfields to break
\sym.  Indeed, we need at least two such superfields: $H$ has
hypercharge $Y=-1/2$, while $\bar{H}$ has $Y=+1/2$. There are at least
three reasons for this. First, we saw in Sec.~3e that a model with a
single Higgs doublet superfield suffers from quadratic divergencies,
since the trace of the hypercharge generator does not vanish. This
already hints at the second problem: A model with a single Higgs
doublet superfield has nonvanishing gauge anomalies associated with
fermion triangle diagrams. The contribution from a complete generation
of SM fermions does vanish, of course, since the SM is anomaly--free.
However, if we only add a single higgsino doublet, anomalies will be
introduced; we need a second higgsino doublet with opposite
hypercharge to cancel the contribution from the first doublet.

Finally, as discussed in Sec.~3f, the masses of chiral fermions must be 
supersymmetric, i.e. they must originate from terms in the superpotential.
On the other hand, we have seen in Sec.~3d that the superpotential must not
contain products of left--chiral and right--chiral superfields. This means that
we are not allowed to introduce the hermitean conjugate of a Higgs superfield
(or of any other chiral superfield) in $f$. It would then be impossible to 
introduce $U(1)_Y$ invariant terms that give masses to both up--type and
down--type quarks if there is only one Higgs superfield; we again need (at
least) two doublets.

Having specified the field content of the MSSM, we have to define the 
interactions. Of course, the gauge interactions are determined uniquely by
the choice of gauge group, which we take to be $SU(3) \times \sym$ as in the
SM. However, while the gauge symmetries constrain the superpotential $f$,
they do not fix it completely. We can therefore appeal to a principle of
minimality and only introduce those terms in $f$ that are {\em necessary}
to build a realistic model. Alternatively, we can demand that $f$ respects
lepton and baryon number; these are automatic (``accidental'') symmetries
of the SM, but could easily be broken explicitly in the MSSM, as we will see
below. Either approach leads to the following superpotential:
\be \label{e58}
f_{\rm MSSM} = \sum_{i,j=1}^3 \left[ \left( \lambda_E \right)_{ij} H L_i
E^c_j + \left( \lambda_D \right)_{ij} H Q_i D^c_j + \left( \lambda_U
\right)_{ij} \bar{H} Q_i U^c_j \right] + \mu H \bar{H}.
\ee
Here $i$ and $j$ are generation indices, and contractions over $SU(2)$ and
$SU(3)$ indices are understood. For example,
\ben \label{e59} \beq
H \bar{H} &\equiv H_1 \bar{H}_2 - H_2 \bar{H}_1; \label{e59a} \\
Q D_R^c &\equiv \sum_{n=1}^3 Q_n \left( D_R^c \right)_n. \label{e59b}
\eeq
\een
The matrices $\lambda_D$ and $\lambda_U$ give rise to quark masses and to
the mixing between quark current eigenstates as described by the familiar
KM matrix \cite{21}. Since the superpotential (\ref{e58}) leaves neutrinos
exactly massless, as in the SM, the matrix $\lambda_E$ can be taken to be
diagonal.

The choice of the superpotential (\ref{e58}) leads to ``R parity conservation''.
Note that the gauge interactions described by eqs.(\ref{e40}), (\ref{e42}) and
(\ref{e42}) only introduce terms in the Lagrangian that contain an even
number of superpartners (gauginos, sfermions or higgsinos). For example,
if $\Phi$ is a matter superfield, the first term in eq.(\ref{e40}) has two
sparticles (sfermions); the second, none; the third, four (sfermions, after
eq.(\ref{e43}) has been used); and the fourth, two (one sfermion and one
gaugino). If $\Phi$ is a Higgs superfield, the first term in eq.(\ref{e40})
contains no sparticles; the second, two (higgsinos); the third, none; and
the fourth, two (one higgsino and one gaugino). The fact that gauge interactions
always involve an even number of sparticles implies that they respect an
``R parity'', under which all SM fields (matter fermions, and Higgs and gauge
bosons) are even while all sparticles (sfermions, higgsinos and gauginos) are
odd.

The interactions produced by the superpotential (\ref{e58}) also
respect this symmetry; this can easily be verified by plugging it into
eq.(\ref{e39}). This means that in the MSSM one has to produce
sparticles in pairs, if one starts with beams of ordinary particles.
For example, one can produce a pair of sleptons from the decay of a
(virtual) $Z$ boson using the first term in eq.(\ref{e40}). Since we
saw in Sec.~3f that sparticles have to be quite heavy, this constraint
reduces the ``mass reach'' of a given collider for sparticle searches.
For example, at $e^+e^-$ colliders one can generally only produce
sparticles with mass below the beam energy, which is only half the
total center--of--mass energy.

Furthermore, a sparticle can only decay into an odd number of other
sparticles and any number of SM particles. For example, a squark might
decay into a quark and a higgsino via a Yukawa interaction described
by the first term of eq.(\ref{e39}), if this decay is kinematically
allowed. In the MSSM the lightest supersymmetric particle (LSP)
therefore cannot decay at all; it is absolutely stable. This gives
rise to characteristic signatures for sparticle production events at
colliders, which allow to distinguish such events from ``ordinary'' SM
events \cite{23}. The argument goes as follows. Since LSPs are stable,
some of them must have survived from the Big Bang era. If LSPs had strong
or electromagnetic interactions, many or most of these cosmological relics
would have bound to nuclei. Since the LSPs would have to be quite massive
in such scenarios, this would give rise to ``exotic isotopes'', nuclei with
very strange mass to charge ratios. Searches \cite{24} for such exotics have
led to very stringent bounds on their abundance, which exclude all models
with stable charged or strongly interacting particles unless their mass exceeds
several TeV \cite{25}. In the context of the MSSM this means that the LSP must
be neutral. As far as collider experiments are concerned, an LSP will then
look like a heavy neutrino, that is, it will not be detected at all, and will
carry away some energy and momentum. Since {\em all} sparticles will
rapidly decay into (at least) one LSP and any number of SM particles, the MSSM
predicts that {\em each} SUSY event has some ``missing (transverse) 
energy/momentum''.

Note that this property is not an automatic consequence of our choice of
field content and gauge group. We could have introduced the following terms
in the superpotential which explicitly break R parity:
\be \label{e60}
f_{\rm R-breaking} = \lambda L L E^c + \lambda' L Q D^c + \lambda'' D^c D^c
U^c + \mu' H L, 
\ee 
where generation indices have been suppressed. The first two terms in
eq.(\ref{e60}) break lepton number L, while the third terms break baryon
number B. Within the MSSM field content one can therefore break R parity
only if either L or B are not conserved; however, in general SUSY models this
relation between B and L conservation on the one hand and R parity on the
other does not hold. If both B and L were broken, the proton would decay very
rapidly; at least some of the couplings in eq.(\ref{e60}) therefore have to
be zero (or very, very small). This makes it very difficult to embed the MSSM
into some Grand Unified model, unless {\em all} the couplings in
eq.(\ref{e60}) are (almost) zero, which we will assume from now on. See
refs.\cite{23} and \cite{26} for further discussions of the theory and
phenomenology of models where R parity is broken.

So far we have only specified the supersymmetry conserving part of the 
Lagrangian. In the gauge and Yukawa sectors we have had to introduce the same
number of free parameters as in the SM; in the Higgs sector the single
parameter $\mu$ replaces two parameters of the SM. However, in general we 
have to introduce a very large number of free parameters to describe SUSY
breaking, as discussed in Sec.~3f:
\begin{eqnarray} \label{e61}
-{\cal L}_{\rm MSSM, \ non-SUSY} &=& m^2_{\tilde q} \left| \tql \right|^2
+ m^2_{\tilde u} \left| \tur \right|^2 + m^2_{\tilde d} \left| \tdr \right|^2
+ m^2_{\tilde l} \left| \tll \right|^2 + m^2_{\tilde e} \left| \ter \right|^2
\nonumber \\
&+& \left( \lambda_E A_E H \tll \ter + \lambda_D A_D H \tql \tdr
+ \lambda_U A_U \bar{H} \tql \tur + B \mu H \bar{H} + h.c. \right)
\nonumber \\
&+& m_H^2 \left| H \right|^2 + m^2_{\bar H} \left| \bar{H} \right|^2 
+ \frac{1}{2} M_1 \wt{B} \wt{B} + \frac{1}{2} M_2 \wt{W} \wt{W}
+ \frac{1}{2} M_3 \wt{g} \wt{g}.
\end{eqnarray}
Here I have used the same symbols for the Higgs scalars $H, \ \bar{H}$ as for
the corresponding superfields. Note that $m^2_{\tilde q}, \ m^2_{\tilde u}, \ 
m^2_{\tilde d}, \ m^2_{\tilde l}$ and $m^2_{\tilde e}$ are in general
hermitean $3 \times 3$ matrices, while $\lambda_U A_U, \ \lambda_D A_D$
and $\lambda_E A_E$ are general $3 \times 3$ matrices. If we allow these
parameters to be complex, the SUSY breaking piece (\ref{e61}) of the Lagrangian
contains more than 100 unknown real constants! Fortunately most processes will
be sensitive only to some (small) subset of these parameters, at least at
tree level. Finally, note that eq.(\ref{e61}) also respects R parity. 
Introducing R parity breaking terms like \tll \tll \ter\ would lead to an
unstable vacuum, i.e. the scalar potential would be unbounded from below,
unless we also introduce the corresponding terms in the superpotential
(\ref{e60}).

This completes the definition of the MSSM. We are now ready to investigate some
of its properties in more detail.

\subsection*{4b. Electroweak Symmetry Breaking in the MSSM}
Given that the (still hypothetical) existence of elementary Higgs bosons leads
to the main motivation for the introduction of weak--scale supersymmetry, it
seems reasonable to start the discussion of the phenomenology of the MSSM with
a treatment of its Higgs sector. This will also lead to a very strong and in
principle readily testable prediction.

Of course, one wants \sym\ to be broken spontaneously, i.e. the 
scalar potential should have its absolute minimum away from the origin.
Let us for the moment focus on the part of the potential that only depends on
the Higgs fields. It receives three types of contributions: Supersymmetric
``F--terms'' from the last term in eq.(\ref{e39}) only contribute mass terms
$\mu^2 \left( |H|^2 + | \bar{H} |^2 \right)$; supersymmetric ``D--terms'',
eq.(\ref{e44}), give rise to quartic interactions; and the SUSY breaking
part (\ref{e61}) of the Lagrangian gives additional mass and mixing terms.
Altogether, one finds:
\begin{eqnarray} \label{e62}
V_{\rm Higgs} &=& m_1^2 \left| H \right|^2 + m_2^2 \left| \bar{H} \right|^2
+ \left( m_3^2 H \bar{H} + h.c. \right) \nonumber \\
&+& \frac {g_1^2 + g_2^2} {8} \left( \left| H^0 \right|^2 - \left|
\bar{H}^0 \right|^2 \right)^2 + \left( {\rm D-terms \ for \ }H^-, \ \bar{H}^+
\right),
\end{eqnarray}
where $g_1$ and $g_2$ are the $U(1)_Y$ and $SU(2)$ gauge couplings, and the
mass parameters are given by
\ben \label{e63} \beq
m_1^2 &= m^2_H + \mu^2; \label{e63a} \\
m_2^2 &= m^2_{\bar H} + \mu^2 ; \label{e63b} \\
m_3^2 &= B \cdot \mu. \label{e63c} 
\eeq \een

One first has to check that one can still choose the vacuum
expectation values such that charge is conserved in the absolute
minimum of the potential (\ref{e62}). This is indeed the case. By
using $SU(2)$ gauge transformations, one can (e.g.) chose $\langle H^-
\rangle = 0$, without loss of generality.  The derivative $\partial
V_{\rm Higgs} / \partial H^-$ can then only be made to vanish if
$\langle \bar{H}^+ \rangle = 0$ as well.\footnote{This equation has a
second solution, $g^2 H^{0\ast} \bar{H}^{0\ast} = 2m_3^2$; however, it
is easy to see that this does not correspond to a minimum of the
potential.} We can therefore ignore the charged components $H^-, \
\bar{H}^+$ when minimizing the potential.\footnote{Even though the
Higgs sector conserves charge, it might still be broken in the
absolute minimum of the complete potential, where some sfermions may
have nonzero vev. See ref.\cite{casas} for a detailed discussion of
this point.} Furthermore, $v \equiv \langle H^0 \rangle$ and $\bar{v}
\equiv \langle \bar{H}^0 \rangle$ can be chosen to be real. The only
contribution to the potential (\ref{e62}) that is sensitive to the
complex phases of the fields is the term $m_3^2 H^0 \bar{H}^0 + h.c$,
which (for real $m_3^2$) is minimized if ${\rm sign} (v \bar{v} ) = -
{\rm sign} (m_3^2)$.  This means that CP invariance cannot be broken
spontaneously in the MSSM.

Note that the strength of the quartic interactions is determined by the
gauge couplings here; in contrast, in the nonsupersymmetric SM the strength of
the Higgs self--interaction is an unknown free parameter. Moreover, in the
direction $ \left| H^0 \right| = \left| \bar{H}^0 \right|$, the quartic term
in (\ref{e62}) vanishes identically. The potential is therefore only bounded 
from below if
\be \label{e64}
m_1^2 + m_2^2 \geq 2 \left| m_3^2 \right|.
\ee
This condition implies that $m_1^2$ and $m_2^2$ cannot both be negative. 
Nevertheless we can still ensure that the origin of the potential is only a
saddle point, i.e. that \sym\ is broken in the minimum of the potential,
by demanding that the determinant of second derivatives of the potential
(\ref{e62}) at the origin is negative, which requires
\be \label{e65}
m_1^2 m_2^2 < m_3^4.
\ee
It is important to note that the conditions (\ref{e64}) and (\ref{e65}) 
cannot be satisfied simultaneously if $m_1^2 = m_2^2$. Further, 
eqs.(\ref{e63}a,b) show that the supersymmetric contribution to $m_1^2$
and $m_2^2$ is the same; any difference between these two quantities must be
due to the SUSY breaking contributions $m^2_H$ and $m^2_{\bar H}$. In other
words, in the MSSM there is an intimate connection between gauge symmetry
breaking and SUSY breaking: The former is {\em not possible} without the latter.

The Higgs potential can now be minimized straightforwardly by solving the
equations $\partial V_{\rm Higgs} / \partial H^0 = \partial V_{\rm Higgs} /
\partial \bar{H}^0 = 0$. Usually it is most convenient to solve these
equations for some of the parameters in eq.(\ref{e62}), rather than for the
vevs $v$ and $\bar{v}$. The reason is that the combination of vevs
\be \label{e66}
\frac {g_1^2 + g_2^2} {2} \left( v^2 + \bar{v}^2 \right) = M_Z^2
= (91.18 \ {\rm GeV})^2
\ee
is very well known. We can therefore describe both vevs in terms of a single
parameter,
\be \label{e67}
\tanb \equiv \bar{v} / v.
\ee
The minimization conditions can then be written as
\ben \label{e68} \beq
m_1^2 &= -m_3^2 \tanb - \frac {1}{2} M_Z^2 \cos(2\beta) ; \label{e68a} \\
m_2^2 &= -m_3^2 \cot \! \beta + \frac{1}{2} M_Z^2 \cos(2\beta). \label{e68b}
\eeq \een
This form is most convenient for the calculation of the Higgs mass matrices
described below. Alternatively, one can use eqs.(\ref{e63}) to derive
\ben \label{e69} \beq
B \cdot \mu &= \frac{1}{2} \left[ \left( m^2_H - m^2_{\bar H} \right)
\tan(2\beta) + M_Z^2 \sin(2\beta) \right] ; \label{e69a} \\
\mu^2 &= \frac {m^2_{\bar H} \sin^2 \beta - m^2_H \cos^2 \beta}
{\cos(2\beta)} - \frac{1}{2} M_Z^2. \label{e69b}
\eeq \een
This form is most convenient if one has some (predictive) ansatz for the
soft SUSY breaking terms; one such example will be discussed in Sec.~4d.

After symmetry breaking, three of the eight degrees of freedom
contained in the two complex doublets $H$ and $\bar{H}$ get ``eaten''
by the longitudinal modes of the $W^\pm$ and $Z$ gauge bosons. The
five physical degrees of freedom that remain form a neutral
pseudoscalar Higgs bosons $A_p$, two neutral scalars $h_p$ and $H_p$,
and a charged Higgs boson $H_p^\pm$, where the subscript $p$ stands
for ``physical''; this is to be compared with the single physical
neutral scalar Higgs boson of the SM. The tree--level mass matrices
for these Higgs states can most easily be computed from the matrix of
second derivatives of the Higgs potential (\ref{e62}), taken at its
absolute minimum. The physical pseudoscalar Higgs boson $A_p$ is made
from the imaginary parts of $H^0$ and $\bar{H}^0$, which have the mass
matrix [in the basis $(\Im H^0/\sqrt{2},\Im \bar{H}^0/\sqrt{2})$]:
\be \label{e70}
{\cal M}^2_I = \mbox{$ \left( \begin{array}{cc}
- m_3^2 \tanb & -m_3^2 \\
- m_3^2 & - m_3^2 \cotb 
\end{array} \right), $}
\ee
where I have used eqs.(\ref{e68}). Note that $\det {\cal M}^2_I = 0$; the
corresponding massless mode is nothing but the neutral would--be Goldstone
boson $G^0 = \frac{1} {\sqrt{2}} \left( \cosb \Im H^0 - \sinb \Im \bar{H}^0
\right)$. The physical pseudoscalar is orthogonal to $G^0$:
$A_p = \frac{1}{\sqrt{2}} \left( \sinb \Im H^0 + \cosb \Im \bar{H}^0
\right)$, with mass
\be \label{e71}
m^2_A = {\rm tr} {\cal M}^2_I = - \frac {2 m_3^2} {\sin (2 \beta)}.
\ee
[Recall that sign$(v \bar{v}) \equiv {\rm sign} (\sin 2 \beta) = - 
{\rm sign}(m_3^2)$.]

Note that $m_A^2 \rightarrow 0$ as $m_3^2 \rightarrow 0$. Such a massless
pseudoscalar ``axion'' is excluded experimentally (if it is connected to
\sym\ breaking; other, ``invisible'' axions are still allowed \cite{27}).
This massless state occurs since for $m_3^2 = 0$, the Higgs potential
(\ref{e62}) is invariant under an additional global $U(1)$, where both $H$
and $\bar{H}$ have the same charge. This new global symmetry is also broken by
the vevs, giving rise to an additional Goldstone boson. The $m_3^2$ term
breaks this symmetry explicitly, thereby avoiding the existence of an axion.
Furthermore, eq.(\ref{e69a}) shows that $m_3^2 = 0$ implies $\sin (2\beta) = 0$,
i.e. $v \cdot \bar{v} = 0$. This means that a nonvanishing $m_3^2$ is also
necessary to give vevs to both Higgs bosons, which in turn are needed to 
give masses to both up--type and down--type quarks, as can be seen from
eq.(\ref{e58}).

This causes an (aesthetic) problem. We have seen above that in the MSSM,
\sym\ breaking requires SUSY breaking. Now we find that we can break \sym\
in a phenomenologically acceptable way only if $m_3^2 \neq 0$, which
implies $\mu \neq 0$, see eq.(\ref{e63c}). This means that we need to
introduce mass parameters both in the supersymmetry breaking and
in the supersymmetry conserving parts of the Lagrangian. Moreover, the two
kinds of dimensionful parameters must be of roughly the same order of
magnitude. Such a connection between these two sectors of the theory is, at
this level at least, quite mysterious. However, several solutions to this
``$\mu-$problem'' have been suggested; see ref.\cite{26} for a further 
discussion of this point.

The neutral scalar Higgs bosons are mixtures of the real parts of $H^0$ and
$\bar{H}^0$. The relevant mass matrix is in the basis 
$(\Re H^0/\sqrt{2},\Re \bar{H}^0/\sqrt{2})$:
\be \label{e72}
{\cal M}^2_R = \mbox{$ \left( \begin{array}{cc}
- m_3^2 \tanb + M_Z^2 \cos^2 \beta & m_3^2 - \frac{1}{2} M_Z^2 \sin(2\beta) \\
m_3^2 -\frac{1}{2} M_Z^2 \sin(2\beta) & - m_3^2 \cotb + M_Z^2 \sin^2 \beta 
\end{array} \right). $}
\ee
Note that $\det {\cal M}^2_R = m_A^2 M_Z^2 \cos^2(2\beta)$ goes to zero if
either $m_A \rightarrow 0$, or $M_Z \rightarrow 0$, or $\tanb \rightarrow 1$
[which implies $\cos(2\beta) \rightarrow 0$]. These three different limits
therefore all lead to the existence of a massless Higgs boson (at least at
tree--level). In general, the eigenvalues of ${\cal M}^2_R$ are given
by:
\be \label{e73}
m^2_{H,h} = \frac{1}{2} \left[ m_A^2 + M_Z^2 \pm \sqrt{ \left( m_A^2 + M_Z^2
\right)^2 - 4 m_A^2 M_Z^2 \cos^2(2\beta) } \right].
\ee
This leads to the important upper bound \cite{28}
\be \label{e74}
m_h \leq \min(m_A, M_Z) \cdot |\cos(2\beta)|.
\ee
The MSSM seems to predict that one of the neutral Higgs scalars must be
lighter than the $Z$ boson! The origin of this strong bound can be traced back
to the fact that the only Higgs self couplings in eq.(\ref{e62}) are electroweak
gauge couplings. In contrast, in the nonsupersymmetric SM the strength of this
coupling is unknown, and no comparable bound on the Higgs mass can be derived.

Unfortunately the bound (\ref{e74}) receives radiative corrections already at
the 1--loop level, the dominant contribution coming from top--stop loops
\cite{29}. These become large if stop masses are significantly bigger than
$m_t$. At scales between the stop and top masses, the Higgs sector should more
properly be described as in the non--supersymmetric SM, where the top 
Yukawa coupling gives a sizable correction to the quartic Higgs self coupling,
and hence to the Higgs mass. In leading logarithmic approximation the bound
(\ref{e74}) is then modified to
\be \label{e75}
m^2_h \leq M_Z^2 \cos^2(2\beta) + \frac {3 m_t^4} {32 \pi^2 \sin^2 \beta
M_W^2} \log \frac {m_{\tilde{t}_1} m_{\tilde{t}_2} } {m_t^2},
\ee
where $m_{\tilde{t}_1}$ and $m_{\tilde{t}_2}$ are the masses of the two stop
eigenstates (see Sec.~4c). Numerically this gives
\be \label{e76}
m_h \leq 130 \ {\rm GeV},
\ee
if one assumes that stop masses do not exceed 1 TeV significantly, and uses
the bound\footnote{Note that the relevant top mass in eq.(\ref{e75}) is the
running $\overline{\rm MS}$ mass taken at scale $\sqrt{m_t \cdot m_{\tilde t}}$
\cite{30}, which is some 10 GeV smaller than the pole mass $m_t({\rm pole}) =
175 \pm 6$ GeV \cite{31}.} $m_t < 185$ GeV; the upper limit in (\ref{e76})
includes a (rather generous) contribution of about 10 GeV from non--logarithmic
corrections \cite{30}.\footnote{The alert reader might wonder why the form of
the correction (\ref{e75}) is so different from the corrections
we computed in Sec.~2, eq.(\ref{e9}). The reason is that eq.(\ref{e75}) has
been derived by requiring that the vevs remain fixed. The corrections
$\propto m^2_{\tilde t}$ to the Higgs mass parameter $m_2^2$ that we found 
in Sec.~2 are therefore absorbed by a change of the tree--level value of
that parameter. Of course, at some point this will lead to unacceptable
finetuning; this is why one usually does not consider stop masses (greatly)
exceeding 1 TeV. After this procedure, the dominant corrections to the
{\em physical} Higgs masses only grow logarithmically with the stop masses,
as indicated in eq.(\ref{e75}).}

The prediction (\ref{e76}) is in principle quite easily testable. If the bound
(\ref{e75}) is (nearly) saturated, $h_p$ becomes very similar to the Higgs boson
of the SM \cite{32}. In particular, the $ZZh_p$ coupling becomes maximal in this
limit. One can then detect the production of $h_p$ in the process
\be \label{e77}
e^+ e^- \longrightarrow Z h_p
\ee
If $m_h$ falls well below the bound (\ref{e75}), the $ZZh_p$ coupling might be
very small, in which case the rate for reaction (\ref{e77}) becomes too small
to be useful. However, one can then look for
\be \label{e78}
e^+ e^- \longrightarrow h_p A_p \ \ \ \ \ \ {\rm or} \ \ \ \ \ \
e^+ e^- \longrightarrow Z H_p.
\ee
By searching for reactions (\ref{e77}) and (\ref{e78}) together, one can
cover the entire parameter space of the MSSM \cite{33}, {\em provided} one has
an $e^+ e^-$ collider with center--of--mass energy $\sqrt{s} \geq 300$ GeV.
No such collider exists as yet; however, there are plans in various countries
to build a linear $e^+ e^-$ collider with $\sqrt{s} \geq 500$ GeV \cite{33a}.
If experiments at such a collider fail to find at least one Higgs boson, the
MSSM can be completely excluded, independent of the values of its 100 or so
free parameters.

In fact, an only slightly weaker version of this statement holds in {\em all}
models with weak--scale supersymmetry, {\em if} one requires that all couplings
of the theory remain in the perturbative regime, i.e. {\em if} the theory
remains weakly coupled, up to some very high energy scale of order
of the GUT scale $M_X \simeq 10^{16}$ GeV. In such more general models, which
introduce new Higgs self couplings by introducing Higgs singlets, the upper
bound (\ref{e76}) could increase to something like 150 GeV \cite{34}; it
remains true, however, that an $e^+ e^-$ collider with $\sqrt{s} \geq 300$
GeV has to discover at least one Higgs boson \cite{34a}.

The last physical Higgs boson of the MSSM is the charged $H_p^\pm$, with mass
\be \label{e79}
m^2_{H^\pm} = M_W^2 + m_A^2;
\ee
notice that it is always heavier than the $W$ boson.\footnote{This is
not necessarily true in more general SUSY models \cite{34}.} The case
$m_A^2 \gg M_Z^2$ is of particular interest. In this ``decoupling
limit'' $A_p, \ H_p$ and $H_p^\pm$ are all very close in mass; they
essentially form a degenerate $SU(2)$ doublet. Furthermore, $m_h$ is
close to its upper bound and, as mentioned earlier, the couplings of
$h_p$ approach those of the single Higgs boson of the SM. Since in
this scenario $h_p$ would be the only Higgs boson that can be
discovered at the next round of colliders \cite{32}, it would be
difficult to distinguish between the SM and the MSSM by just studying the
Higgs sector. However, one could still deduce that the SM should cease to
describe Nature at a relatively low energy scale, beyond which ``new physics''
of some sort has to appear. The reason is that in the SM the renormalization
group running of the quartic Higgs coupling would lead to the scalar potential
becoming unbounded from below at a scale $\Lambda \ll M_X$, if the Higgs mass is
below 150 GeV or so \cite{35}. Searches for Higgs bosons therefore play a very
important role in testing SUSY in general and the MSSM in particular. So far,
the most stringent bounds on the Higgs sector come from experiments at the
$e^+ e^-$ collider LEP \cite{21}:
\ben \label{e80} \beq
m_h &\geq 62 \ {\rm GeV}, \hspace*{2cm} {\rm if} \ m_A^2 \gg M_Z^2;
\label{e80a} \\
m_h, \ m_A &\geq 45 \ {\rm GeV}, \hspace*{2cm} {\rm if} \ m_A \simeq m_h.
\label{e80b}
\eeq \een

This concludes my discussion of \sym\ breaking in the MSSM. More detailed
studies can be found e.g. in refs.\cite{32}.

\subsection*{4c. Sparticle Mixing}
Once \sym\ is broken, fields with different \sym\ quantum numbers can mix,
if they have the same $SU(3)_c \times U(1)_{\rm em}$ quantum numbers. The
Dirac masses of the SM quarks and leptons can be understood as such mixing
terms, since they couple a left--handed $SU(2)$ doublet to a right--handed
singlet. A closely related phenomenon occurs in the sfermion sector of the
MSSM.

All three types of contributions to the scalar potential (F--terms, D--terms
and SUSY breaking terms) appear in the sfermion mass matrices. I will for the
moment ignore mixing between sfermions of different generations, but will 
include mixing between $SU(2)$ doublet and singlet sfermions. The sfermion
mass matrix then decomposes into a series of $2 \times 2$ matrices, each of
which describes sfermions of a given flavor. Let us consider the case of the
scalar top. The F--term contribution (\ref{e39}) gives rise to diagonal
\ttl\ and \ttr\ masses from $\left| \partial f / \partial \ttr \right|^2$ and
$\left| \partial f / \partial \ttl \right|^2$, respectively; these contributions
are equal to $m_t^2$. F--terms also give an off--diagonal contribution from
$\left| \partial f / \partial \bar{H}^0 \right|^2$, which is proportional to
$\lambda_t v \mu = m_t \mu \cotb$. The D--terms (\ref{e44}) only give rise to
diagonal mass terms, but these differ for \ttl\ and \ttr, since these fields
transform differently under \sym. Finally, the soft breaking terms (\ref{e61})
give (in general different) contributions to the diagonal \ttl\ and \ttr\
mass terms, as well as an off--diagonal contribution $\propto A_t m_t$. 
Altogether one has \cite{36} [in the basis (\ttl, \ttr)]:
\be \label{e81}
{\cal M}^2_{\tilde t} = \mbox{$ \left( \begin{array}{cc}
m_t^2 + m^2_{\tilde{t}_L} + \left( \frac{1}{2} - \frac{2}{3} \stw \right)
\dtm & - m_t \left( A_t + \mu \cotb \right) \\
- m_t \left( A_t + \mu \cotb \right) & m_t^2 + m^2_{\tilde{t}_R} + 
\frac{2}{3} \stw \dtm
\end{array} \right). $}
\ee

Note that the off--diagonal entries are $\propto m_t$. Eq.(\ref{e81}) also
describes the $\tilde{c}$ and $\tilde{u}$ mass matrices, with the obvious
replacement $t \rightarrow c$ or $u$. Since $m_c, \ m_u \ll m_{\tilde c}, \
m_{\tilde u}$, $\tilde{u}_L - \tilde{u}_R$ and $\tilde{c}_L - \tilde{c}_R$
mixing are usually negligible.\footnote{An exception can occur for (loop)
processes where chirality arguments imply that {\em all} contributions are
suppressed by small quark masses; in such cases the off--diagonal entries in
the $\tilde{u}$ and $\tilde{c}$ mass matrices must be included.} However, since
the top mass is comparable to the other masses that appear in eq.(\ref{e81}),
\ttl--\ttr\ mixing is generally important. To mention but one example, even
though the $Z \ttl \tilde{t}_L^\ast$ and $Z \ttr \tilde{t}_R^\ast$ couplings are
nonzero, the coupling of the $Z$ boson to a physical stop eigenstate of the
matrix (\ref{e81}), $\tilde{t}_1 = \ttl \cos \! \theta_{\tilde t} +
\ttr \sin \! \theta_{\tilde t}$, vanishes \cite{37} if $\cos^2 
\theta_{\tilde t} = \frac{4}{3} \stw$.

The calculation of the sbottom mass matrix is completely analogous to that of
${\cal M}^2_{\tilde t}$. Ohe has [in the basis (\tbl, \tbr)]:
\be \label{e82}
{\cal M}^2_{\tilde d} = \mbox{$ \left( \begin{array}{cc}
m_b^2 + m^2_{\tilde{t}_L} - \left( \frac{1}{2} - \frac{1}{3} \stw \right)
\dtm & - m_b \left( A_b + \mu \tanb \right) \\
- m_b \left( A_b + \mu \tanb \right) & m_b^2 + m^2_{\tilde{b}_R} - 
\frac{1}{3} \stw \dtm
\end{array} \right). $}
\ee
Note that the soft breaking mass that appears in the $(1,1)$ entry of 
${\cal M}^2_{\tilde b}$ is the same as that in the $(1,1)$ entry of 
${\cal M}^2_{\tilde t}$. This is a consequence of $SU(2)$ invariance: If the
SUSY breaking piece (\ref{e61}) of the Lagrangian contained different masses
for members of the same doublet, $SU(2)$ would be broken explicitly and the
theory would no longer be unitary on the quantum level. This leads to the 
important relation
\be \label{e83}
m^2_{\tilde{l}_L} = m^2_{\tilde{\nu}_l} - M_W^2 \cos(2\beta),
\ee
which holds if $\tilde{l}_L - \tilde{l}_R$ mixing can be neglected; this is
always the case for $l=e$ or $\mu$. However, in general no such relation
holds for masses of $SU(2)$ singlet sfermions.

The off--diagonal entries of the matrix (\ref{e82}) are again proportional to
the relevant quark mass. Nevertheless, $\tbl-\tbr$ (as well as 
$\tilde{\tau}_L - \tilde{\tau}_R$) mixing can be important \cite{38} if
$\tanb \gg 1$. Such scenarios are viable, since the $b$ quark is the heaviest
fermion that gets its mass from the vev $v$, which therefore need not be
larger than a few GeV. Note that the top mass is $\propto \bar{v}$. This
implies that the top and bottom Yukawa couplings will be close to each other
if $\tanb \simeq m_t/ m_b \simeq 50$; this is necessary in certain Grand
Unified models based on the group $SO(10)$ \cite{39}. Notice, however, that
\tbl--\tbr\ mixing, if it is important at all, is driven by $\mu$, while the
dominant contribution to \ttl--\ttr\ mixing usually comes from $A_t$.
($A_{b,t} \gg m_{\tilde b, \tilde t}$ is forbidden, since it leads to a 
charge and colour breaking absolute minimum of the scalar potential 
\cite{frere}.)

Mixing between \ttl\ and \ttr\ and, if $\tanb \gg 1$, between \tbl\
and \tbr\ as well as between $\tilde{\tau}_L$ and $\tilde{\tau}_R$ is
quite generic. In contrast, mixing between sfermions of different
generations is very model dependent. Such mixing can cause severe
phenomenological problems, by producing unacceptably large flavor
changing neutral currents (FCNC) between ordinary quarks and leptons
through 1--loop processes. Such problems occur if the mass matrix for
squarks with a given charge does not commute with the corresponding
quark mass matrix, since then there will be flavor off--diagonal
gluino--quark--squark couplings. This can easily be seen by writing
the physical (mass) eigenstates as $\left( q_p \right)_i = \sum_j
\left( U_q \right)_{ij} q_j$ and $\left( \tilde{q}_p \right)_i =
\sum_j \left( U_{\tilde q} \right)_{ij}
\tilde{q}_j$, where $q_j$ and $\tilde{q}_j$ are current eigenstates. This
gives [see eq.(\ref{e40})]:
\begin{eqnarray} \label{e84}
{\cal L}_{\tilde{g} \tilde{q} q} & \propto & \overline{\tilde g} \sum_{i=1}^{3} 
q_i \tilde{q}_i^\ast + h.c \nonumber \\
&=& \overline{\tilde g} \sum_{i=1}^3 \left[ \sum_{j=1}^3 \left( U_q^\dagger
\right)_{ij} \left( q_p \right)_j \right] \cdot \left[ \sum_{l=1}^3 \left(
U^T_{\tilde q} \right)_{il} \left( \tilde{q}^\ast_p \right)_l \right] + h.c.
\nonumber \\
&=& \overline{\tilde g} \sum_{j,l=1}^3 \left( U_{\tilde q} U^\dagger_q
\right)_{lj} \left( q_p \right)_j \left( \tilde{q}^\ast_p \right)_l + h.c.
\end{eqnarray}
This will be flavor--diagonal only if the matrices $U_q$ and $U_{\tilde q}$
can be chosen to be equal, which is possible if the $q$ and $\tilde{q}$ mass
matrices commute. This condition is trivially satisfied if squarks of a given
charge all have the same mass, in which case their mass matrix is proportional
to the unit matrix, but other possibilities also exist. A recent analysis
\cite{40} finds that constraints on $K^0 - \overline{K^0}$ and 
$D^0 - \overline{D^0}$ mixing force the off--diagonal entries in eq.(\ref{e84})
between the first and second generation to be very small, unless squarks and
gluinos are significantly heavier than 1 TeV. The corresponding bounds
involving third generation (s)quarks are somewhat weaker.

Similar problems also arise in the slepton sector, if one replaces the gluino
in eq.(\ref{e84}) with a bino or neutral wino. In this case the most severe
constraints come from $\mu \rightarrow e \gamma $ decays and $\mu \rightarrow
e$ conversion in muonic atoms. Finally, chargino loop contributions to
$K^0 - \overline{K^0}$ mixing limit mass splitting between $SU(2)$ doublet
squarks of the first and second generation, (almost) independently of any
mixing angles \cite{41}.

The breaking of \sym\ also leads to mixing between electroweak gauginos and
higgsinos. This mixing is caused by the last term in eq.(\ref{e40}), which
can couple a Higgs boson to a gaugino and a higgsino; when the Higgs field is
replaced by its vev, these terms give rise to off--diagonal entries in the
``chargino'' and ``neutralino'' mass matrices. The physical charginos 
$\tc^+_{1,2}$ are therefore mixtures of the charged $SU(2)$ gauginos and the
charged higgsinos. Their mass matrix in the (gaugino, higgsino) basis can be
written as
\be \label{e85}
{\cal M}_\pm = \mbox{$ \left( \begin{array}{cc}
M_2 & \sqrt{2} M_W \sin \! \beta \\
\sqrt{2} M_W \cos \! \beta & \mu
\end{array} \right). $}
\ee
Notice that ${\cal M}_\pm$ is not symmetric, unless $\tanb=1$. In general one
therefore needs two different diagonalization matrices for the right-- and
left--handed components of the charginos; see the first paper in ref.\cite{32}
for a careful discussion of this point.

The neutralinos are mixtures of the $\wt{B}$, the neutral $\wt{W}$, and the
two neutral higgsinos. In general these states form four distinct Majorana
fermions, which are eigenstates of the symmetric mass matrix [in the basis
$(\wt{B}, \ \wt{W}, \ \tilde{h}^0, \ \tilde{\bar h}^0 )$]:
\be \label{e86}
{\cal M}_0 = \mbox{$ \left( \begin{array}{cccc}
M_1 & 0 & - M_Z \cos \! \beta \sin \! \theta_W & M_Z \sin \! \beta \sin \!
\theta_W \\
0 & M_2 &   M_Z \cos \! \beta \cos \! \theta_W & -M_Z \sin \! \beta \cos \!
\theta_W \\
- M_Z \cos \! \beta \sin \! \theta_W & M_Z \cos \! \beta \cos \! \theta_W & 
0 & -\mu \\
M_Z \sin \! \beta \sin \! \theta_W & -M_Z \sin \! \beta \cos \! \theta_W &
-\mu & 0
\end{array} \right). $}
\ee
The masses and mixing angles of the charginos and neutralinos are therefore
completely determined by the values of the four parameters $M_1, \ M_2, \ \mu$
and \tanb. In most analyses one further reduces the dimensionality of this
parameter space by assuming that gaugino masses unify at the GUT scale 
$M_X \simeq 10^{16}$ GeV. This is motivated by the observation that in the MSSM
the three gauge couplings do seem to meet at this scale, if one uses their
experimentally determined values as inputs at scale $M_Z$ and ``runs'' them to
higher scales using their renormalization group equations (RGE) \cite{41a}.
The gaugino masses $M_i$ run in the same way as the corresponding squared
gauge couplings $g_i^2$ do. Assuming $M_1 = M_2$ at scale $M_X$ then
implies
\be \label{e87}
M_1(M_Z) = \frac{5}{3} \tan^2 \theta_W M_2(M_Z) \simeq \frac{1}{2} M_2(M_Z),
\ee
where the factor 5/3 comes from the difference between the GUT normalization
and the usual SM normalization of the hypercharge generator.

If eq.(\ref{e87}) holds, the phenomenology of the charginos and neutralinos
is essentially fixed by three parameters.\footnote{Their decay branching
ratios in general also depend on sfermion and Higgs masses, however.} If
$|\mu| > |M_2| \geq M_Z$, the two lightest neutralino states will be
dominated by the gaugino components, with $\tc^0_1$ being mostly $\wt{B}$ and
$\tc^0_2$ being mostly $\wt{W}^0$; similarly, the light chargino $\tc^\pm_1$
will be mostly a charged wino. In this case one very roughly finds 
$m_{\tc_1^\pm} \simeq m_{\tc_2^0} \simeq 2 m_{\tc_1^0}$. In the opposite
limit, $|\mu| < |M_1|$, the two lighter neutralinos and the lighter chargino
are all mostly higgsinos, with masses close to $|\mu|$. Finally, if $|\mu|
\simeq |M_2|$, some of the states will be strongly mixed. The size of the
mixing also depends to some extend on \tanb. If \tanb\ is not large,
there will be considerably more mixing if $M_2 \cdot \mu \cdot \tanb >
0$ than for the opposite choice of sign; notice that for this sign,
the two terms in the determinant of the chargino mass matrix tend to
cancel, and a similar cancellation occurs in det${\cal M}_0$. Such mixing
lowers the mass of the light eigenstates, and increases that of the heavy
ones.

The details of the chargino and neutralino sectors are of importance for many
areas of MSSM phenomenology. For example, $\tc_1^0$ is usually taken to be
the LSP; we saw in Sec.~4a that the LSP must be electrically and color neutral.
This means that any other sparticle will eventually decay into a $\tc_1^0$.
However, in many cases the $\tc_1^0$ is produced only at the end of a lenghty
decay chain or ``cascade'' \cite{42}. Consider the case of the gluino. If
gaugino masses are unified, the gluino mass $|M_3|$ at the weak scale is about
3.5 times larger than $|M_2|$. The gluino will therefore decay into the
lighter electroweak gauginos and a pair of quarks via the exchange of real or
virtual squarks. However, since $g_2^2 \simeq 3.3 g_1^2$, gluinos will prefer
to decay into the $SU(2)$ gaugino states, even though these are heavier than
the $U(1)_Y$ gaugino. If $|\mu| < |M_2|$, gluinos will therefore decay into
the {\em heaviest} chargino and neutralinos, which in turn will decay by the
exchange of gauge or Higgs boson or a sfermion; often these decays will
themselves proceed in several steps. The situation can become even more
complicated if top quarks can be produced in gluino decays, since the (s)top
has large Yukawa couplings to higgsino--like charginos and neutralinos. A
realistic treatment of gluino decays has to keep track of all these different
possibilities. Space does not permit to delve into these details any further
here; I refer the interested reader to ref.\cite{23}, where many more references
can be found.

\subsection*{4d. Minimal Supergravity}
As mentioned at the end of Sec.~4a, a general parametrization of supersymmetry
breaking in the MSSM introduces about 100 free parameters. This is 
not very satisfactory. The predictive power of a theory with such a large number
of parameters is clearly quite limited, although we saw in Sec.~4b that some
interesting predictions can be derived even in this general framework. It
is therefore desirable to try and construct models that make do with fewer
free parameters. The oldest and, to my mind, still most elegant such model
goes under the name minimal supergravity (mSUGRA).

This model is based on the {\em local} version of supersymmetry.
Eqs.(\ref{e12b}) show that invariance under local SUSY transformations
implies invariance under local coordinate change; this invariance is
the principle underlying Einstein's construction of the theory of
General Relativity. Local supersymmetry therefore naturally includes
gravity, and is usually called supergravity.

We saw in Sec.~2f that it is quite difficult to break global supersymmetry
spontaneously, partly because this necessarily creates a cosmological constant.
This is no longer true for supergravity. To see that, one first introduces the
``K\"ahler potential''
\be \label{e88}
G = - \sum_i X_i (\phi_j) \phi_i \phi_i^\ast - M^2_{Pl} \log \frac
{\left| f(\phi_j) \right|^2} {M^6_{Pl}},
\ee
where $f$ is again the superpotential, and the $X_i$ are real
functions of the chiral fields $\phi_j$. Supersymmetry is broken
spontaneously if $\langle G_i \rangle\equiv \langle \partial G /
\partial \phi_i \rangle \neq 0$ for some $i$; in the ``flat limit''
$\mpl \rightarrow \infty$ this reduces to $\langle
\partial f / \partial \phi_i \rangle \neq 0$, which is the condition for
F--term breaking of global supersymmetry, see eq.(\ref{e56}). For
finite \mpl, the F--term contribution to the scalar potential becomes
\be \label{e89}
V_F = M^2_{Pl} e^{-G/M^2_{Pl}} \left[ G^i \left( G^{-1} \right)^j_i G_j +
3 M^2_{Pl} \right],
\ee
where $G^i \equiv \partial G / \partial \phi^\ast_i$, and $\left( G^{-1} 
\right)^j_i$ is the matrix of second derivatives of $G$. There can be a
cancellation between the two terms in the square bracket in eq.(\ref{e89});
one can therefore simultaneously have $\langle G_i \rangle \neq 0$ (broken
SUSY) and $\langle V_F \rangle = 0$ (vanishing cosmological constant).

It is nevertheless still very difficult to break supersymmetry using
only the fields present in the MSSM. One therefore introduces a
``hidden sector'' \cite{42a}, which consists of some fields that do
not have any gauge or superpotential couplings to the ``visible
sector'' containing the MSSM. Nevertheless, the supergravity
Lagrangian \cite{6a} {\em automatically} transmits SUSY breaking from
the hidden to the visible sector, through operators that are
suppressed by some powers of \mpl. In the simplest models, the order of
magnitude of the soft breaking terms (\ref{e61}) will be set by the gravitino
mass,
\be \label{e90}
m_{3/2} = \mpl e^{-G/(2 M^2_{Pl})} = \frac {| \langle f \rangle|}
{M^2_{Pl}} \exp \left( \sum_i X_i \phi_i \phi^\ast_i / M^2_{Pl} \right).
\ee
The most natural choice is to give some hidden sector field(s) vev(s) of order
\mpl. Then $\langle f \rangle \sim {\cal O} ( \mpl \langle f_i \rangle)$,
and the second factor in eq.(\ref{e90}) is of order unity. Requiring the
soft breaking masses to be $\sim {\cal O}(M_Z)$ then implies
\be \label{e91}
\langle f_i \rangle \sim {\cal O}(M_Z \cdot \mpl) \sim \left (10^{10} \
{\rm GeV} \right)^2.
\ee
However, one can also construct models where the soft breaking masses in the
visible sector are either much smaller or much larger than the gravitino
mass \cite{grav}.\footnote{Formally \cite{6a}, the ``SUSY breaking scale''
$M_S^2 = - \mpl e^{- \langle G \rangle / (2 M^2_{Pl})} \left \langle 
(G^{-1})^i_j G_i \right\rangle$. Note that $(G^{-1})^i_j = - \delta^i_j$
for fields with canonical kinetic energy terms. This gives
$
M_S^2 = e^{-X \phi^2/M^2_{Pl}} \left[ \frac {\langle X \phi_i f \rangle }
{M^2_{Pl}} + \langle f_i \rangle \right].
$
Under the ``natural'' assumptions listed above, this reproduces 
approximately the numerical value given in eq.(\ref{e91}).}

The ansatz (\ref{e88}) is not very predictive; in general it still
gives different soft terms for different scalar fields \cite{43,10}. The
number of free parameters can, however, be reduced dramatically by 
imposing a global $U(N)$ symmetry on the ``K\"ahler metric''
$(G^{-1})^i_j$, where $N$ is
the number of superfields in the visible sector (17 in the MSSM, for three
generations plus two Higgs doublets). This implies in particular that the
functions $X_i$ must be the same for all MSSM fields. One then finds \cite{44}
that SUSY breaking in the visible sector can be described using only three
parameters:
\ben \label{e92} \beq
m_i^2 &= m_0^2 \hspace*{2cm} \forall i \label{e92a} \\
A_{ijk} &= A_0 \hspace*{2cm} \forall i,j,k \label{e92b} \\
B_{ij} &= B_0 \hspace*{2cm} \forall i,j \label{e92c}
\eeq \een
In the MSSM there is in any case only one $B-$parameter; however, 
eqs.(\ref{e92a},b) are very restrictive indeed. Finally, in mSUGRA one
assumes that the gaugino masses are unified at the GUT scale:
\be \label{e93}
M_1(M_X) = M_2(M_X) = M_3(M_X) = m_{1/2}.
\ee

In principle eqs.(\ref{e92}) should hold at an energy scale close to \mpl,
where gravitational interactions are integrated out. In practice one usually
assumes that these ``boundary conditions'' still hold at the GUT scale
$M_X$. This is not a bad approximation if there is no full GUT field
theory, e.g. if the MSSM directly merges into superstring theory (which
also predicts unification of gauge couplings). However, if a GUT exists,
eqs.(\ref{e92}) will in general receive sizable corrections \cite{45}.
Unfortunately these corrections depend quite sensitively on the poorly known
details of the GUT sector.

Even if eqs.(\ref{e92}) still hold to good approximation at scale $M_X$,
the spectrum at the weak scale is significantly more complicated. The
reason is that the soft breaking parameters ``run'', i.e. depend on the
energy scale, just like the gauge couplings do. Indeed, we already saw in
the previous subsection that the gaugino masses have the same scale
dependence as the squared gauge couplings; eq.(\ref{e93}) therefore
implies
\be \label{e94}
M_a(Q) = m_{1/2} \frac {g_a^2(Q)} {g_a^2(M_X)},
\ee
so that $M_3 : M_2 : M_1 \simeq 7 : 2 : 1$ at the weak scale.

Scalar masses also receive corrections from gauge interactions, involving
both gauge boson loops and gaugino--fermion (or, in case of Higgs bosons,
gaugino--higgsino) loops, as discussed in Sec.~2e. In fact, each class
of contributions by itself is quadratically divergent, but these divergencies
cancel between the bosonic and fermionic contributions. However, if SUSY is
broken, a logarithmic divergence remains, which contributes to the running
of the scalar masses:\footnote{Note that the logarithmic divergencies we
found in Sec.~3e should be absorbed in the running of the $SU(2)$ gauge
coupling, not the running of the Higgs mass.}
\be \label{e95}
\left. \frac {d m^2_i} {d \log Q } \right|_{\rm gauge} = - \sum_{a=1}^{3}
\frac {g_a^2}{8 \pi^2} c_a(i) M_a^2,
\ee
where $a$ labels the gauge group, and $c_a(i)$ is a group factor; e.g.
$c_3(\tilde{q}) = 16/3$ for $SU(3)$ triplets, $c_2(\tilde{l}_L) = 3$ for
$SU(2)$ doublets, and $c_1(\phi) = 4 Y_{\phi}^2$ for fields with
hypercharge $Y_{\phi}$. Note the negative sign in eq.(\ref{e95}); it
implies that gauge interactions will {\em increase} scalar masses as
the scale $Q$ is reduced from $M_X$ to the weak scale. Obviously the
sfermions with the strongest gauge interactions will receive the
largest corrections. This leads to the predictions
\ben \label{e96} \beq
1 & \leq m_{\tilde q}/ m_{\tilde{l}_L} \leq 3.5 ; \label{e96a} \\
1 & \leq m_{\tilde{l}_L} /  m_{\tilde{l}_R} \leq 1.9 \label{e96b},
\eeq \een
where the lower bounds hold for $m_{1/2} \rightarrow 0$ and the upper
bounds for $m_0 \rightarrow 0$. Note that eqs.(\ref{e96}) hold only if
Yukawa interactions are negligible, which is true for the first two
generations, as well as for the stau's and $\tilde{b}_R$ unless 
$\tanb \gg 1$.

The effect of Yukawa interactions on the running of Higgs masses has
almost been computed in Sec.~2, eq.(\ref{e11}); the only missing piece
is a contribution from the wave function renormalization of the Higgs
field, which is proportional to the Higgs mass itself. Altogether, one
finds for the $\bar{H}$ doublet \cite{46}:
\be \label{e97}
\frac {d m^2_{\bar H}} {d \log Q} = \frac {3 \lambda_t^2} {8 \pi^2}
\left( m^2_{\bar H} + m^2_{\tilde{t}_L}  + m^2_{\tilde{t}_R} +
A_t^2 \right) + ( {\rm gauge \ terms}).
\ee
Since the second Higgs doublet, $H$, only has Yukawa interactions
involving $\lambda_b$ and $\lambda_{\tau}$, radiative corrections
from Yukawa interactions give {\em different} contributions to
$m_H^2$ and $m^2_{\bar H}$. This is important, since we saw in Sec.~4b,
eqs.(\ref{e64}) and (\ref{e65}), that successful \sym\ breaking requires
$m_H^2 \neq m^2_{\bar H}$ at the weak scale, whereas the boundary condition
(\ref{e92a}) stipulates their equality at the Planck or GUT scale. Note also
that the sign in eq.(\ref{e97}) is positive. This means that Yukawa 
interactions will {\em reduce} $m^2_{\bar H}$ as we go down from $M_X$ to
$M_Z$. In fact, given that $\lambda_t \geq 1$ at the weak scale, 
$m^2_{\bar H}$ can quite easily become negative at scales $Q$ exponentially
smaller than $M_X$, thereby triggering \sym\ breaking \cite{46,47}. This is
called ``radiative symmetry breaking'', since radiative corrections as
described by eq.(\ref{e97}) play a crucial role here.\footnote{One
occasionally sees the statement in the literature that $\lambda_t \geq 1$,
i.e. a heavy top quark, is necessary for this mechanism to work. This is not
true, since condition (\ref{e65}) can be satisfied even if both $m^2_H$ and
$m^2_{\bar H}$ are positive. Indeed, for historical reasons there is an
extensive literature \cite{48} on radiative symmetry breaking with a top mass 
around 40 GeV. It is true, however, that the allowed parameter space opens up
as $\lambda_t$ is increased.}

The stop mass parameters $m^2_{\tilde{t}_R}$ and $m^2_{\tilde{t}_L}$  
also receive corrections $\propto \lambda_t^2$. However, the group (color)
factor 3 in eq.(\ref{e97}) is reduced to 2 and 1, respectively, since for
stops the color index is fixed on the external legs;  the correction to
$m^2_{\tilde{t}_R}$ receives a factor of two due to summation over $SU(2)$
indices in the loop. The group structure of the MSSM therefore naturally 
singles out the Higgs fields, since their masses are reduced most by
radiative corrections involving Yukawa interactions. Notice also
that  $m^2_{\tilde{t}_{L,R}}$ in general receive large positive corrections
from $SU(3)$ gauge interactions, while the gauge terms in eq.(\ref{e97})
only involve the much weaker \sym\ interactions.

Eq.(\ref{e97}) implies that the Higgs mass parameters at the weak
scale depend in a quite complicated fashion on the GUT--scale
parameters $m_0, \ m_{1/2}, \ A_0$ and $B_0$, as well as on the top
Yukawa coupling. Of course, we still want to arrange things such the
$Z$ boson gets its proper mass. One way to ensure this is to chose as
free parameters the set $(m_0, \ m_{1/2}, \ A_0, \ \tanb, \ {\rm
sign}\mu)$, and to use eqs.(\ref{e69}) to determine $B$ and $|\mu|$ at
the weak scale. This is a convenient choice since (at least at
one--loop level) the RGE for $B$ and $\mu$ are decoupled from those
for the other soft breaking parameters \cite{46}. The value of $B$ is
of little interest per se, since this parameter only appears in the
Higgs potential. The parameter $\mu$ also appears in many mass
matrices, however; see eqs.(\ref{e81}), (\ref{e82}), (\ref{e85}) and
(\ref{e86}).  Since $\lambda_t$ is large, $m^2_{\bar H}$ usually
becomes quite negative at the weak scale, and one needs a sizable
$\mu^2$ in eq.(\ref{e69b}); hence one usually (but not always) has
$|\mu| \geq |M_2|$ at the weak scale \cite{38,49}. An exception can
occur for $m^2_0 \gg m^2_{1/2}$, where $|\mu| < m_{1/2}$ is still possible.
Similarly, one finds that three of the four Higgs bosons are usually quite
heavy. In particular, at tree--level the mass of the pseudoscalar boson
is given by \cite{38}
\be \label{e98}
m_A^2 = \frac {m^2_{\tilde \nu} + \mu^2} { \sin^2 \beta} +
{\cal O}(\lambda_b^2, \lambda_{\tau}^2).
\ee
One is thus usually close to the ``decoupling limit'' $m^2_A \gg M^2_Z$
discussed in Sec.~4b. However, the Yukawa terms omitted in eq.(\ref{e98})
are negative; if $\tanb \simeq m_t/m_b$, they can reduce $m_A$ to values at
(or below) its experimental lower bound \cite{38}. In fact, the requirement
$m^2_A > 0$ often determines the upper bound on \tanb.

Space does not allow me to extend this rather sketchy introduction to
mSUGRA phenomenology. The interested reader is referred to refs.\cite{23,26}
for further details (and many additional references). Note also that a
program that implements radiative symmetry breaking starting from the
boundary conditions (\ref{e92}), (\ref{e93}) is part of the ISAJET event
generator \cite{50}; copies of the program are available from 
{\tt baer@hep.fsu.edu}.

\section*{5. Outlook}

Clearly a short introduction to supersymmetry, like the one I have
attempted here, can at best give a flavor of the work that has been,
and is being, done in this very large field. In the main part of these
lectures I had to give short thrift to most recent developments. In
this concluding section I will try to at least briefly sketch some
areas of active research, and provide some of the relevant references.

Recent efforts in SUSY model building, i.e. the construction of
potentially realistic supersymmetric field theories, have mostly
focussed on two quite different approaches. On the one hand, there has
been much interest in models where supersymmetry is broken in a hidden
sector at a rather low scale, and is then mediated to the visible
sector by gauge interactions \cite{15}. One drawback of such models is
that one needs to introduce a new ``messenger sector'' to transmit
SUSY breaking to the MSSM fields; in mSUGRA this is done {\em
automatically} by terms in the Lagrangian whose presence is required
by local supersymmetry. Proponents of this class of models list as its
main advantage that it automatically gives equal masses to sfermion
with equal $SU(3) \times \sym$ quantum numbers, thereby avoiding the
problems with FCNC discussed in Sec.~4b. In fact, however, this is not
true if one writes down the most general superpotential allowed by the
gauge symmetry of the MSSM; one has to introduce additional symmetries
to forbid Yukawa couplings between the messenger and MSSM sectors.

Interest in this class of models peaked a few months ago, since they
seemed to be able to explain a rather strange event \cite{52} observed
by the CDF collaboration at the tevatron $p \bar{p}$ collider, where
the final state consists of an $e^+e^-$ pair and two hard photons; the
transverse momenta of these four particles do not add up to zero, i.e.
there is missing transverse energy.  The probability for this event to
be due to SM processes is estimated to be of the order of $10^{-3}$.
Since in these models the SUSY breaking scale is rather low, the
gravitino $\wt{G}$ (which does not have any gauge interactions) is
very light, see eq.(\ref{e90}). In this case $\tc_1^0 \rightarrow
\wt{G} + \gamma $ decays can have quite short lifetimes \cite{53}, and
the ``CDF event'' could be explained as selectron pair production
followed by $\wt{e} \rightarrow \tc_1^0 + e$ and $\tc_1^0 \rightarrow 
\wt{G} + \gamma$ decays. However, this explanation is beginning to look 
quite unlikely, since one then expects (many) more unusual events in
other final states involving hadronic jets \cite{54}, none of which
seem to have been seen \cite{55}. Also, it was recently pointed out
that at least the simplest models in this class have absolute minima
of the scalar potential where either SUSY remains unbroken in the MSSM
sector, or charge and color are broken \cite{56}. My personal view is
that these models introduce needless complications with no apparent
gain.\footnote{I am a strong believer in Occam's razor, even though it
may seem sexist nowadays.}

In a quite different development, people have been trying to construct
models that describe both the large mass splittings in the SM quark
and lepton sector and the required near--degeneracy of their scalar
superpartners (more accurately, the fact that the sfermion mass
matrices have to almost commute with the fermion mass matrices) by the
{\em same} mechanism. Usually this is achieved by means of a
judiciously chosen discrete \cite{57} or continuous \cite{58}
symmetry. Most of these models have been embedded in some GUT theory,
so whatever new (s)particles they predict (beyond those of the MSSM)
tend to have masses ${\cal O}(M_X)$.  However, they also predict (at
least ``generically'') that sfermion loop contributions to FCNC
processes, while suppressed, are not vanishingly small. This gives new
importance to searches for processes that are forbidden in the SM
(e.g., $\mu \rightarrow e \gamma$ decays), as well as to careful
experimental studies of processes that are allowed but suppressed in
the SM \cite{40} ($b \rightarrow s \gamma$ decays, $K^0 -
\overline{K^0}, D^0 - \overline{D^0}$ and $B^0 - \overline{B^0}$
mixing, various CP--violating asymmetries, ...).

Finally, ``string phenomenologists'' attempt to make contact between superstring
theory and the real world \cite{59}. This approach might well be the most 
promising one in the long run. However, to the best of my knowledge no firm
prediction of phenomenological interest has yet emerged from string theory.

The recent proliferation of SUSY models makes it imperative to devise
ways to distinguish between them. The most direct method is to measure
the masses and couplings of superparticles as accurately as possible.
This has been the focus of many recent studies of SUSY collider
phenomenology \cite{60,60a}. Searching for superparticles at $e^+e^-$
colliders is usually quite straightforward, since the relevant SM
background processes usually have roughly comparable cross--sections;
the presence of massive invisible LSPs in SUSY events then gives them
kinematic properties that allow to distinguish them from SM
backgrounds. However, when one is trying to find ways to measure MSSM
parameters precisely, even small backgrounds can become important.
Furthermore, one may have to distinguish experimentally between
different SUSY processes, which can be much more challenging than
discriminating between SUSY and SM events.

Hunting for sparticles at hadron colliders is more tricky, since now
the cross section for (hard) SM processes often exceeds those for the
SUSY reactions of interest by several orders of magnitude. Another
problem is the relatively more ``messy'' environment in which
experiments at hadron colliders must work. This is partly due to the
fact that the initial state now radiates gluons (rather than photons
at $e^+e-$ colliders), each of which will produce several hadrons.
Another problem is that only some fraction of the $p$ or $\bar{p}$
beam energy goes into the hard (partonic) scattering reaction, with
the rest going into ``beam remnants''. This makes kinematic event
reconstruction much more difficult, since much of the beam remnants
escapes down the beam pipes, carrying an unknown amount of energy and
longitudinal momentum with it. Therefore one often only uses momentum
components transverse to the beam in partial kinematic
reconstructions, and also for the purpose of devising cuts that
increase the signal--to--background ratio.

Nevertheless several promising signals for sparticle production at
hadron colliders have been clearly established in Monte Carlo studies
\cite{23}. Unfortunately searches for these final states at existing
colliders so far have yielded null results; one can conclude from
these studies that (most) squarks and gluinos have to be heavier than
about 200 GeV, the precise bounds depending on details like the squark
to gluino mass ratio \cite{12}. These limits are still well below the
naturalness or finetuning limit of very roughly 1 TeV. However, at the
LHC collider, which is scheduled to commence operations in about a
decade, one should see clear SUSY signals in several different final
states if supersymmetry is to provide a solution to the hierarchy
problem.  Work on how to measure MSSM parameters and distinguish
between competing SUSY models at hadron collider experiments has only
begun relatively recently \cite{60a}.

Finally, a considerable amount of work has gone into studies of
implications of supersymmetry for cosmology and astrophysics. The
perhaps most prominent example is ``Dark Matter'' (DM) \cite{62},
which is known to form most of the mass of the Universe. Studies of
Big Bang nucleosynthesis indicate that at least some of the DM must
not be baryonic. The LSP, being absolutely stable if R--parity is
unbroken, has been known for some time to make a good DM particle
candidate \cite{61}. This is now a relatively mature field \cite{62},
but calculations of the density of LSP relics from the Big Bang era,
and studies of how to detect them experimentally, are still being
refined. The first pilot searches \cite{61a} for DM particles already
exclude the possibility that a stable sneutrino could be the LSP.
These experiments were not sensitive enough to probe much of the MSSM
parameter space if the LSP is the lightest neutralino, but this is
expected to change in the next decade or so. Finally, without going
into any detail I mention that superparticles might also play a
crucial role in generating the baryon asymetry of the Universe
\cite{63,64}, and perhaps even in the conjectured very early ``inflationary''
stage of the development of the Universe \cite{64}.

It should be clear by now that supersymmetrists have penetrated almost
all areas of particle physics research. You are welcome to join us!

\subsection*{Acknowledgements}
I thank Xerxes Tata for a careful reading of this manuscript.
I thank the Center for Theoretical Physics, Seoul National University,
as well the the KEK Theory Group, for their hospitality while I was
writing these notes.

\end{document}